\documentclass[times,5p]{elsarticle}

\usepackage[T1]{fontenc}
\usepackage{microtype}
\usepackage{lmodern}
\usepackage{booktabs}
\usepackage[dvipsnames]{xcolor}
\usepackage{amsmath,amssymb,bm}
\usepackage{xspace}
\usepackage{enumitem}
\usepackage{listings}
\usepackage{lineno}
\usepackage[pdfencoding=unicode,linkcolor=blue,citecolor=blue,urlcolor=blue]{hyperref}
\usepackage[subrefformat=parens,labelfont=bf,font={stretch=1.15}]{subcaption}

\usepackage{cleveref}
\usepackage[version=4]{mhchem}
\usepackage[object=vectorian]{pgfornament}
\usepackage{siunitx}[=v2]

\journal{Astroparticle Physics}

\newcommand{\cor}[1]{CORSIKA~{#1}\xspace}

\DeclareSIUnit\yr{yr}

\newcommand{\threeByTwoGrid}[6]{
    \begin{minipage}[t]{0.03\textwidth}
        \centering
        \raisebox{-2.85cm}{\rotatebox{90}{0.2 rad}}
    \end{minipage}%
    \begin{minipage}[t]{0.32\textwidth}
        \centering
        \textbf{$f_{\mathrm{total}}$} \\
        \includegraphics[width=\textwidth]{#1}
    \end{minipage}%
    \hfill
    \begin{minipage}[t]{0.32\textwidth}
        \centering
        \textbf{$f_{\mathrm{geo}}$} \\
        \includegraphics[width=\textwidth]{#2}
    \end{minipage}%
    \hfill
    \begin{minipage}[t]{0.32\textwidth}
        \centering
        \textbf{$f_{\mathrm{ce}}$} \\
        \includegraphics[width=\textwidth]{#3}
    \end{minipage}
    
    \vspace{1em} 

    \begin{minipage}[t]{0.03\textwidth}
        \centering
        \raisebox{1.8cm}{\rotatebox{90}{0.001 rad}}
    \end{minipage}%
    \begin{minipage}[t]{0.32\textwidth}
        \centering
        \includegraphics[width=\textwidth]{#4}
    \end{minipage}%
    \hfill
    \begin{minipage}[t]{0.32\textwidth}
        \centering
        \includegraphics[width=\textwidth]{#5}
    \end{minipage}%
    \hfill
    \begin{minipage}[t]{0.32\textwidth}
        \centering
        \includegraphics[width=\textwidth]{#6}
    \end{minipage}
}

\begin{document}
\begin{frontmatter}
\title{Simulating radio emission from particle cascades with \cor8}
\cortext[cor1]{Email address: corsika8@kit.edu}
\author[1,2]{J.M. Alameddine}
\author[1,2]{J. Albrecht}
\author[3]{J. Ammerman-Yebra}
\author[4]{L. Arrabito}
\author[5,6]{A.A. Alves Jr.}
\author[1,2]{D. Baack}
\author[7]{A. Coleman}
\author[1,2]{\mbox{H. Dembinski}} 
\author[1,2]{D. Elsässer}
\author[5]{R. Engel}
\author[4]{A. Faure}
\author[5]{A. Ferrari}
\author[8]{C. Gaudu}
\author[7]{C. Glaser}
\author[5]{M. Gottowik}
\author[5]{D. Heck}
\author[5,9]{\mbox{T. Huege}} 
\author[8]{K.H. Kampert}
\author[5]{N. Karastathis}
\author[10]{L. Nellen}
\author[5]{T. Pierog}
\author[11]{R. Prechelt}
\author[12]{M. Reininghaus}
\author[1,2]{W. Rhode}
\author[13,3]{\mbox{F. Riehn}} 
\author[1,2]{M. Sackel}
\author[5]{P. Sampathkumar}
\author[8]{A. Sandrock}
\author[1,2]{J. Soedingrekso}
\author[5]{R. Ulrich}
\affiliation[1]{organization={Technische Universität Dortmund (TU), Department of Physics}, city={Dortmund}, country={Germany}}
\affiliation[2]{organization={Lamarr Institute for Machine Learning and Artificial Intelligence}, city={Dortmund}, country={Germany}}
\affiliation[3]{organization={Universidade de Santiago de Compostela, Instituto Galego de Física de Altas Enerxías (IGFAE)}, city={Santiago de Compostela}, country={Spain}}
\affiliation[4]{organization={Laboratoire Univers et Particules de Montpellier, Université de Montpellier}, city={Montpellier}, country={France}}
\affiliation[5]{organization={Karlsruhe Institute of Technology (KIT), Institute for Astroparticle Physics (IAP)}, city={Karlsruhe}, country={Germany}}
\affiliation[6]{organization={University of Cincinnati, Cincinnati}, city={OH}, country={United States}}
\affiliation[7]{organization={Uppsala University, Department of Physics and Astronomy}, city={Uppsala}, country={Sweden}}
\affiliation[8]{organization={Bergische Universität Wuppertal, Department of Physics}, city={Wuppertal}, country={Germany}}
\affiliation[9]{organization={Vrije Universiteit Brussel, Astrophysical Institute}, city={Brussels}, country={Belgium}}
\affiliation[10]{organization={Universidad Nacional Autónoma de México (UNAM), Instituto de Ciencias Nucleares}, city={Ciudad de México}, country={México}}
\affiliation[11]{organization={University of Hawai'i at Manoa, Department of Physics and Astronomy}, city={Honolulu}, country={USA}}
\affiliation[12]{organization={Karlsruhe Institute of Technology (KIT), Institute of Experimental Particle Physics (ETP)}, city={Karlsruhe}, country={Germany}}
\affiliation[13]{organization={Laboratório de Instrumentação e Física Experimental de Partículas (LIP)}, city={Lisboa}, country={Portugal}}

\begin{abstract}
CORSIKA 8 is a new framework for simulations of particle cascades in air and dense media implemented in modern C++17, based on past experience with existing codes, in particular CORSIKA~7. The flexible and modular structure of the project allows the development of independent modules that can produce a fully customizable particle shower simulation. The radio module in particular is designed to treat the electric field calculation and its propagation through complex media to each observer location in an autonomous and flexible way. It already allows for the simultaneous simulation of the radio emission calculated with two independent time-domain formalisms, the ``Endpoint formalism'' as previously implemented in CoREAS and the ``ZHS'' algorithm as ported from ZHAireS. The design acts as the baseline interface for current and future development for the simulation of radio emission from particle showers in standard and complex scenarios, such as cross-media showers penetrating from air into ice. In this work, we present the design and implementation of the radio module in CORSIKA~8, along with validation studies and a direct comparison of the radio emission from air showers simulated with CORSIKA~8, CORSIKA~7 and ZHAireS. We also present the impact of simulation details such as the step size of simulated particle tracks on radio-emission simulations and perform a direct comparison of the ``Endpoints'' and ``ZHS'' formalisms for the same underlying air showers. Finally, we present an in-depth comparison of CORSIKA~8 and CORSIKA~7 for optimum simulation settings and discuss the relevance of observed differences in light of reconstruction efforts for the energy and mass of cosmic rays.
\end{abstract}
\begin{keyword}
Monte Carlo simulations, air shower simulations, radio emission
\end{keyword}
\end{frontmatter}


\section{Introduction}
\label{sec:introduction}

When an ultra-high-energy particle impinges on the Earth's atmosphere it creates a cascade of secondary particles, known as an extensive air shower. Studying the characteristics of air showers and their development in the atmosphere provides crucial information on the nature and origin of the primary cosmic particle. The main information of interest are the energy, mass and arrival direction of the primary particle. For this, many experiments have been constructed that utilize a variety of detection techniques with the most popular and well-established ones being the particle and fluorescence light detection techniques ~\cite{KAMPERT2012660}. For the past 20 years, though, due to technological advancements (mainly in the fields of digital signal processing and computing capacity) and a detailed understanding of the emission mechanisms of radio signals from air showers, the radio detection technique has undergone an impressive renaissance ~\cite{Huege:2016veh, Schr_der_2017}. Experiments like LOPES \cite{LOPES:2021ipp}, LOFAR \cite{Schellart_2013} and the Auger Engineering Radio Array \cite{aera} have demonstrated the successful use of the radio detection technique. As a result, further experiments like SKA \cite{Huege_ska} and the upgrade of the Pierre Auger Observatory \cite{Castellina_2019,Huege:2023pfb} will utilize the radio detection technique in the future. In addition, sensitivity to ultra-high-energy neutrinos can be obtained by detecting the radio emission generated by neutrino-induced particle showers in air \cite{grandd,Zeolla:2023khf} or ice \cite{Barwick:2022vqt}, the latter being explored in polar regions through in-ice radio arrays or with antenna-equipped balloons that are flown around Antarctica. The novel capabilities of CORSIKA~8 will allow the simulation of the complex geometries relevant for radio neutrino detectors and will be explored in forthcoming work. 

To analyze and reconstruct the experimental data, the community heavily relies on detailed Monte Carlo simulations of air showers. The gold standard so far has been the CORSIKA code~\cite{Heck:1998vt} originally developed for KASCADE in the 1990s. It still serves the air-shower community as a work horse, fulfilling a vast array of simulation needs. Its monolithic FORTRAN structure, lack of flexibility (e.g., regarding particle showers in dense media), and the requirements of modern experiments have, however, made it clear that the astroparticle physics community needs a modern, flexible and modular new simulation software code. Since 2018, CORSIKA~8 (C8)~\cite{Engel:2018akg}, a modern C++-based simulation framework has been in development whose main design goals are modularity and flexibility, making it easier to upgrade and maintain by a new generation of scientists.

The C8 code can now be considered ``physics-complete'' (see \cite{CORSIKA:2023jyz} and references therein). It provides access to the well-established hadronic interaction models EPOS-LHC~\cite{Pierog:2013ria}, Sibyll~2.3d~\cite{Riehn:2019jet}, and QGSJet-II.04~\cite{PhysRevD.83.014018} at high energies, as well as an experimental inclusion \cite{Reininghaus:2023ctx} of Pythia~8.3 as a high-energy interaction model~\cite{Sjostrand:2021dal} for air-shower simulations. Low-energy hadronic interactions are being simulated with FLUKA~\cite{Ferrari:2005zk,Battistoni:2015epi}. Electromagnetic cascades in CORSIKA~8 are being handled with the PROPOSAL \cite{koehne2013proposal, dunsch_2018_proposal_improvements,Alameddine:2020zyd,Alameddine:2023wrp} code, currently in version 7.6.2; details and a comparison to EGS4 as used in CORSIKA~7 (C7) are provided in reference~\cite{AlexanderICRC2023}. Photohadronic interactions in electromagnetic cascades are fully treated. The total energy-dependent cross section for photohadronic interactions is provided by PROPOSAL, the low-energy treatment of particle production is treated with SOPHIA~\cite{Mucke:1999yb}, and the high-energy treatment of particle production is performed with Sibyll. The Landau-Pomeranchuk-Migdal effect is taken into account and a new particle-thinning approach for electromagnetic cascades has been developed and implemented \cite{CORSIKA:2023jyz}. A dedicated article describing the design and capabilities of C8 in detail is currently in preparation.

Radio emission from air showers has been simulated successfully in the past. On the microscopic level, two formalisms are widely used, namely the ``Endpoints'' formalism~\cite{James:2010vm,Ludwig:2010pf} as implemented in the CoREAS extension \cite{Huege:2013vt} of C7 and the ``ZHS'' formalism~\cite{Alvarez-Muniz:2010wjm} as implemented in ZHAireS~\cite{zhairess}. These formalisms have been derived from first principles to calculate the electric field resulting from each charged particle in the cascade individually. Models that describe the radio emission macroscopically such as MGMR~\cite{Scholten_2008} or EVA~\cite{Werner_2012}, although faster, do not achieve the same level of accuracy and thus microscopic simulations are typically relied upon for analyses. The main focus of this article is the presentation of the radio module in C8, its low-level validation, and a comparison of C8 radio-emission simulations against those of established simulation codes, taking into account details of the underlying particle simulation setup.

This article is structured as follows: in section \ref{sec:physics}, the current knowledge of the radio emission mechanisms along with a short review of the two main microscopic formalisms is briefly summarized. In section \ref{sec:design}, the design of the radio module, its components and its integration into C8 are presented. In section \ref{sec:validation}, the radio module is  validated in terms of low-level tests both disconnected from C8 and tightly integrated into C8. A full-blown comparison of the radio emission predicted by C8 for a high-energy iron-induced air shower with the predictions by C7 and ZHAireS is shown in section \ref{sec:air-shower-comparison}. In section \ref{sec:coreas-vs-zhs}, the predictions by the ``Endpoints'' and ``ZHS'' formalisms are directly compared for the same underlying air showers. In section \ref{sec:c8-vs-c7}, a comparison between radio-emission simulations with C8 and C7 for different values of the step size in simulated particle tracks is presented. Finally, in section \ref{sec:c7vsc7} we condlude with an illustration of the sensitivity of radio-emission simulations with C7 to the length of the particle tracks in the underlying shower simulation.

\section{Radio emission from air showers}
\label{sec:physics}

Radio emission from extensive air showers is by now well-understood~\cite{Huege:2016veh}. There are two main mechanisms, so-called geomagnetic emission and charge-excess (or Askaryan) radiation. The geomagnetic effect arises as the Earth's magnetic field deflects electrons and positrons in the shower in opposite directions, inducing a time-varying transverse current which emits radiation polarized linearly along the direction of the Lorentz force. The charge-excess emission arises because of electrons being ionized from atmospheric molecules and being accumulated in the shower front as the shower evolves, leading to a longitudinal time-varying current. The charge-excess emission is linearly polarized along the radial direction, i.e., the orientation of the electric field vector depends on observer position. Both mechanisms contribute to the radio emission, with the geomagnetic emission being dominant in air and the charge-excess emission being vastly dominant in dense media, like ice.

Moreover, due to the refractive index of the media that the radio wave is propagating through, Cherenkov-like effects are observed. At the Cherenkov angle, the shower particles and the radio waves propagate at almost identical velocity and hence the radiation reaching an observer along this viewing angle is compressed in time and thus coherent up to high frequencies of several GHz. In air, where the refractive index is close to unity, the Cherenkov angle is of order $\simeq 1^{\circ}$ which corresponds to distances of $\mathcal{O}$(100\,m) from the shower axis for vertical showers. This creates a \textit{Cherenkov ring} in the radio-emission footprint where the energy deposited at the ground in terms of radio waves is at its maximum.

Recently, it has been found that the radio emission from very inclined air showers exhibits additional features not present for more vertical air showers \cite{Chiche:2024yos}. Especially in the presence of strong magnetic fields, a loss of coherence and the appearance of a ``clover leaf'' pattern in the polarization component perpendicular to both the shower axis and the Lorentz force become visible. The latter had previously only been predicted to be visible at high frequencies \cite{Huege:2013yra} and is typically interpreted as the presence of synchrotron radiation~\cite{Huege_2005}.

In this work we focus on calculations using the microscopic ``Endpoints'' and ``ZHS'' formalisms, which we have ported over from their implementations within C7 and ZHAireS, respectively. In a discretized Monte Carlo simulation of a particle cascade, in particular within the radio-emission calculations, charged particles are treated as propagating with constant velocity on straight track segments approximating their actual curved trajectories. In the ``ZHS'' formalism, electromagnetic radiation is interpreted as coming from the midpoints of these track segments. The vector potential that describes the radiation emitted from such a charged-particle track is calculated in the Fraunhofer approximation, i.e., assuming that both the wavelength and the length of the track are small with respect to the distance from source to observer. Depending on the observer location and track length, the track is thus sub-divided into smaller tracks before the vector potential calculation takes place. The ``Endpoints'' formalism is based on the concept that a charged particle radiates during acceleration. Hence, particles radiate at the start and the end of the tracks where they are accelerated and decelerated, respectively. For observer positions very near the Cherenkov angle, the ``Endpoint'' formalism diverges numerically, whereas the ``ZHS'' formalism can be evaluated analytically for this limiting case. In the implementation of the ``Endpoints'' formalism both in the CoREAS code of C7 \cite{Huege:2013vt} and its C8 version, the ``Endpoints'' calculation thus falls back to a ``ZHS''-like solution for situations where the numerical stability would otherwise be problematic (in air, this occurs when the ``boost factor'' representing the compression of signal arrival time differences versus signal emission time differences, governing the denominator in the endpoints formalism, exceeds the value of 1,000). For clarity, we will refer to this approach as the ``CoREAS'' formalism in the following.

For an in-depth discussion of the conceptual differences between the two formalisms, we refer the interested reader to reference \cite{James:2022mea}. Previous comparisons for the radio emission from particle showers in air and dense media calculated using the two formalisms have shown them to agree well~\cite{Zilles:2015fpr,PierreAuger:2016vya,Gottowik:2017wio}. Yet for the case of air showers the comparisons could so far not disentangle the effects resulting from differences in the formalisms from those of the underlying air-shower simulation. In this work, we showcase the power of our radio module by directly comparing the two formalisms, for the first time, for the same underlying air showers in section \ref{sec:coreas-vs-zhs}, thereby taking out any uncertainty due to the underlying simulation codes and the general impossibility to reproduce two exact same showers with two different codes.

\section{Design of the radio module} 
\label{sec:design}

\begin{figure*}[ht]
 \centering
 \includegraphics[width=0.75\textwidth]{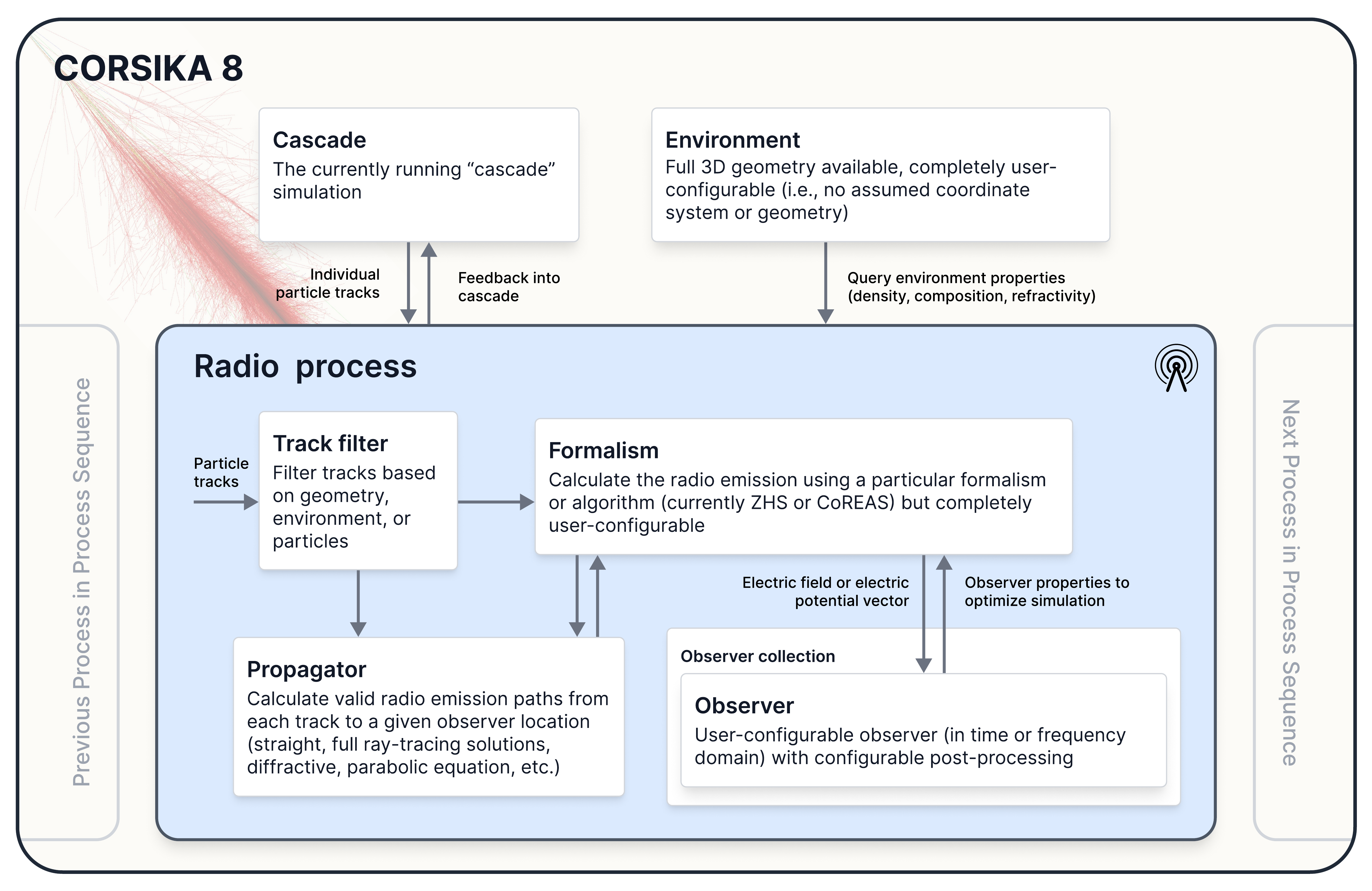}
 \caption{A schematic diagram of the radio process currently implemented in C8, and showing how it integrates with the C8 framework. See text for further explanations.} 
 \label{fig:design}
\end{figure*}

C8 is comprised of (physical) processes taking place within a user-defined geometrical space with properties like density, refractive index, etc., characterizing the interaction media or \textit{Environment}. The particle travels on tracks inside this geometrical space in a manner dictated by the tracking algorithm. All the (physical) processes defined and used for the simulation make up the \textit{Process Sequence} and can affect the trajectories and the lifetime of the particles. The main algorithm of the simulation is the \textit{Cascade} which handles the processing of the particles on the stack from first interaction until the last according to the user-defined \textit{Process Sequence}. For further details we refer the interested reader to reference \cite{Reininghaus:2022qyl}.

C8's structure is highly modular, flexible and easily extensible. The radio module was designed with the same philosophy for it to be able to support current and next-generation radio-detection experiments. The top-level architecture of the radio module is shown in~\cref{fig:design}. When constructed, the radio module initiates a \textit{Radio Process} that retrieves the relevant information for the radio-emission calculation from C8 through the \textit{Cascade} and the user-configured \textit{Environment}. From the \textit{Cascade}, the main loop of the overall simulation, all the particle tracks are fed into the \textit{Radio Process} and from the \textit{Environment} all necessary properties (e.g., refractive index, density, etc. for any given position) are queried.

The radio process is built from separate components implemented as template parameters to allow flexibility for the kind and the number of \textit{Radio Processes} one can instantiate in the same shower simulation. Each component is configurable by the user and can be easily swapped out with either C8-provided or user-provided C++ code. These are:

\paragraph{Track Filter} This is a top level check responsible for selecting which of the particle tracks are forwarded further into the radio calculation. For instance, as radio emission arises almost purely from electrons and positrons in the cascade, the module can dismiss the rest of the particles. Furthermore, it is also possible to specify ``slices'' in terms of particle energy, atmospheric depth or other parameters and calculate the radio emission only from these ``slices'', a feature that turned out to be very useful in the CoREAS extension of C7.

\paragraph{Formalism} This provides the core calculation that the radio module performs. The formalism receives the relevant information from the particle tracks (in particular particle positions and associated times) and applies the appropriate algorithm to calculate the electric field vector at the chosen observer location. Currently, there are two formalisms that are fully implemented, both in the time domain, namely ``CoREAS'' and ``ZHS''. We have ported both of them as closely as possible from their original versions in order to be able to directly compare C8 with C7 and ZHAireS. The structure of the code allows for an easy inclusion of other formalisms in the module, for example frequency-domain versions of these or other formalisms.

\paragraph{Propagator} The \textit{Propagator} is responsible for the signal propagation from the position where a signal is emitted to the desired observer position, along potentially multiple paths through the given \textit{Environment}. For the time being there are three fully implemented propagators, all of them considering the emission as propagating on straight paths. First, we have implemented a propagator which integrates the travel time along the propagation path with a user-defined stepsize, i.e., is computationally expensive, but is able to work in arbitrarily complex media -- as long as straight-ray propagation is a good approximation. Second, we have implemented two propagators for the particular case of atmospheres that can be considered ``flat'' (an approximation good up to air-shower zenith angles of $\approx 65^\circ$) coupled with a uniform-density or exponential-density-gradient atmosphere. The first calculates the signal propagation analytically, while the second uses tables for higher computation speed. In all the propagators provided so far, the refractivity in the atmosphere is coupled to the air density by a proportionality relation, the Gladstone-Dale law~\cite{GladstoneDale}.

The advantage of decoupling the radio-signal propagation from the formalism is that it allows the treatment of more complex scenarios such as cross-media showers~\cite{juan-icrc23,DeKockere:2024qmc} and ray-traced propagation of signals on curved paths \cite{Glaser:2019cws,VandenBroeck:2023aee}. In C8, it will be sufficient to develop a new propagator for the specific experimental needs while using the rest of the radio module as is. 

\paragraph{Observer} The \emph{Observer} object is in charge of processing and storing the calculated electric fields at the desired observer locations in the simulation run. Multiple instances of independent \emph{Observers} can be initialized in the simulation for every \textit{Radio Process}. These make up an \textit{Observer Collection}. The interface allows for added features by the user in order to apply their own unique processing to the received electric field, e.g., they could apply an antenna response on-the-fly. We currently implement a standard time-domain \emph{Observer} that has a sample rate, detection start time, and a time window that is configurable individually for each \emph{Observer}. Frequency-domain \emph{Observers} can also be easily implemented once frequency-domain formalisms become available. 


In summary, the way the \textit{Radio Process} works internally is the following: It receives a particle track from the \textit{Cascade}, then proceeds to filtering and forwarding the track to the \textit{Formalism} in case it is relevant. The \textit{Formalism} loops over the \textit{Observer Collection} and treats each \textit{Observer} separately. At this stage, the \textit{Formalism} calls the \textit{Propagator} to retrieve the propagation path, including the acquired time delay, to the \textit{Observer}. Then the electric field contribution is calculated and, along with its detection time, is provided to the \textit{Observer} for storage. A version of the radio module that takes advantage of multithreading in order to calculate the radio emission for multiple antennas in parallel is also available. The corresponding design, implementation details, and performance gain measurements have been discussed in reference \cite{AugustoICRC2023}. The implementation is in the final stages of development before its integration in the main-line code.

In the next section, we leverage the C8 functionality to apply multiple \textit{Radio Processes} to the same air shower for an in-depth comparison of the predictions of the ``CoREAS'' and ``ZHS'' formalisms. In the future, one-to-one comparisons using different propagators or time-domain versus frequency-domain studies on identical showers could also be an interesting application of this capability.

\section{Validation of the radio module}
\label{sec:validation}

\begin{figure*}
    \centering
    \includegraphics[width=0.49\textwidth]{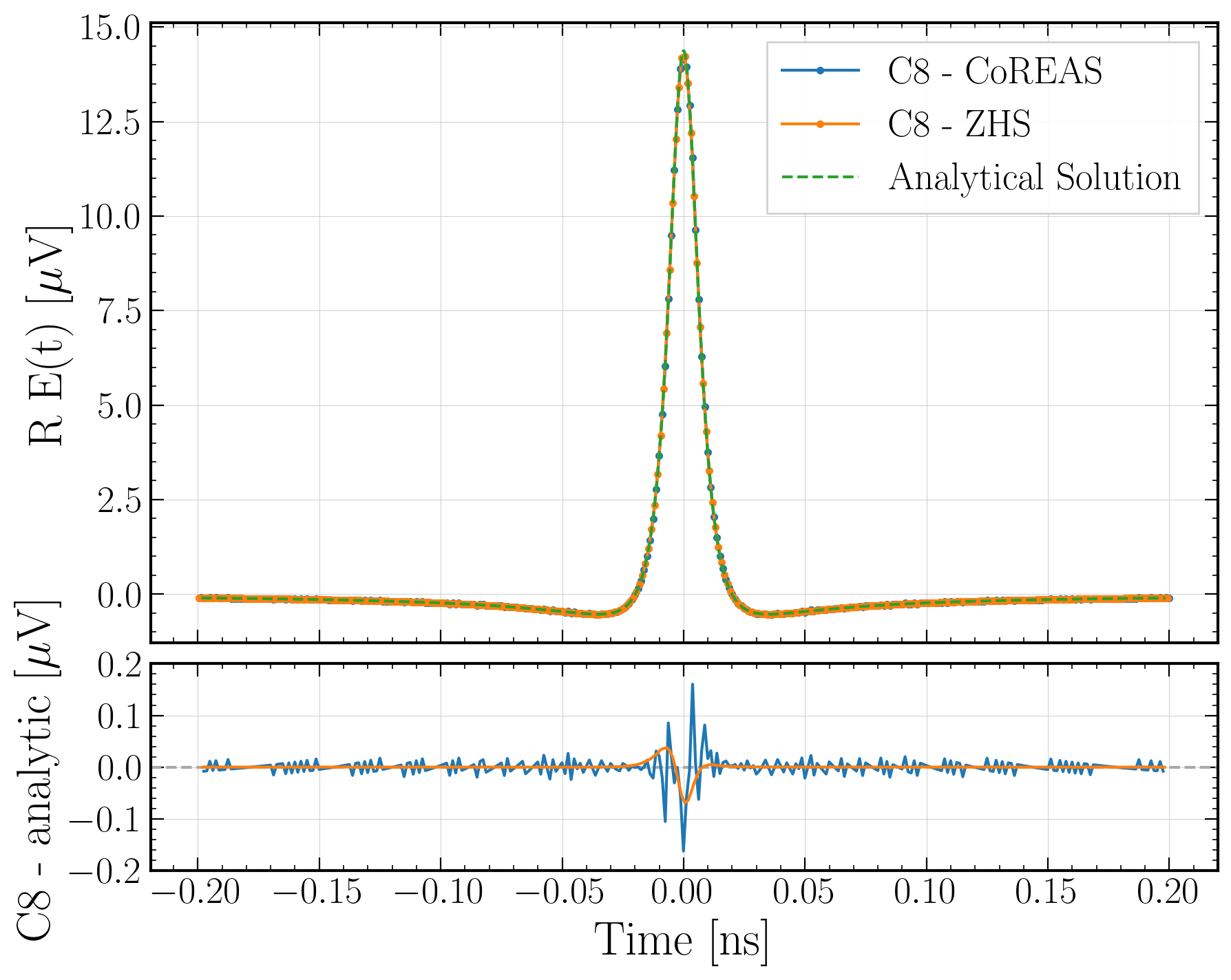}
    \includegraphics[width=0.49\textwidth]{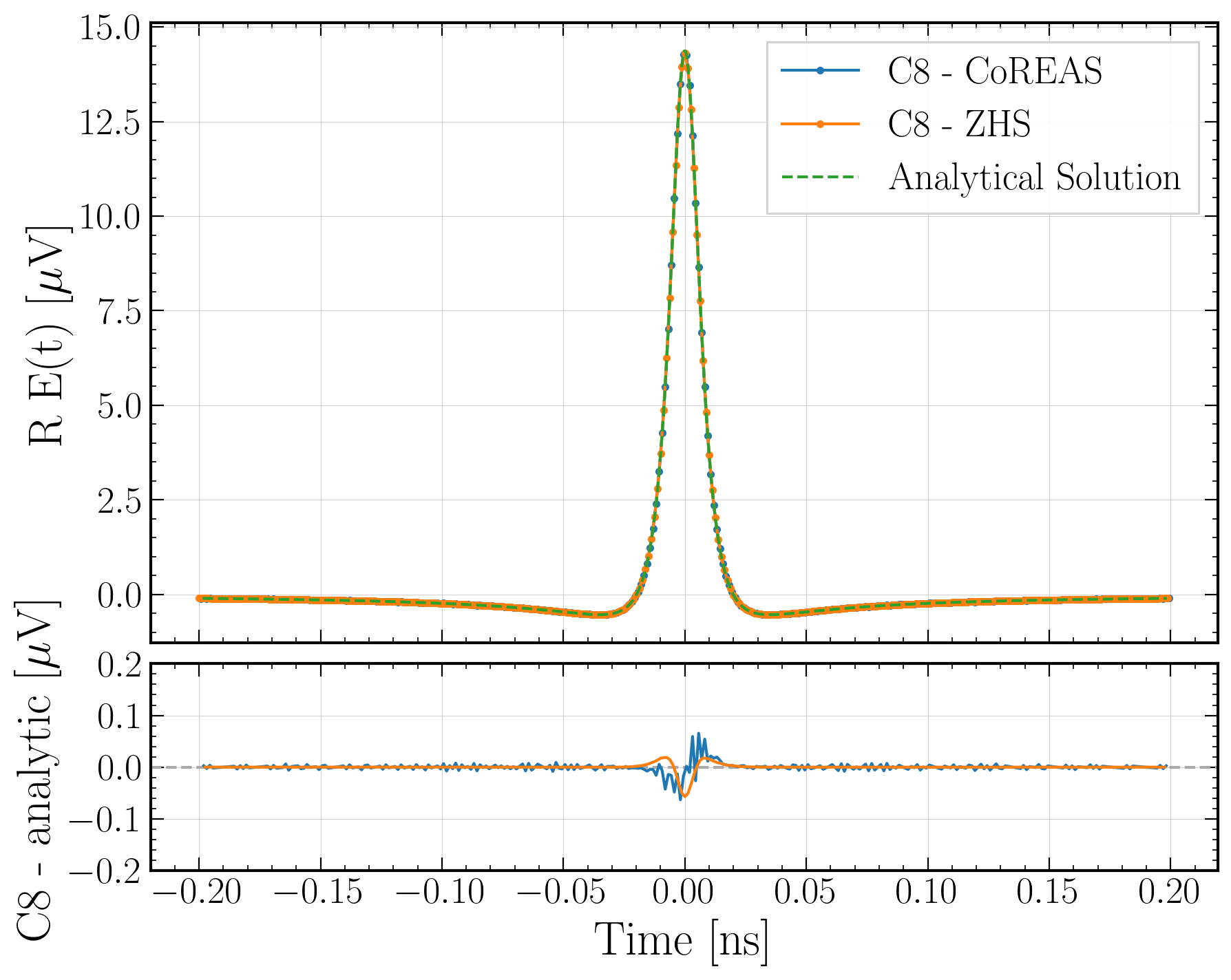}
    \caption{Radio pulse from an electron in a uniform magnetic field performing a circular loop. The y-axis denotes the product of distance $R$ and electric field component $E(t)$. ``Manual tracking'' algorithm (left) and ``C8 tracking'' (right). Both tracking algorithms are compared with the analytical solution presented in~\cite{James:2010vm}.}
    \label{fig:synchall}
\end{figure*}

C8 poses development guidelines characterized by high standards for all its modules and radio is no exception. All classes and methods have to pass extensive unit testing with very high code coverage. In addition to this low-level software testing, in this section we present a validation based on calculations for a physical scenario.

To this end, we consider a relativistic electron in a uniform magnetic field undergoing circular motion in vacuum (refractive index $n=1$). We use both the ``CoREAS'' and ``ZHS'' formalisms to calculate the associated synchrotron emission, and compare the resulting pulses to an analytical solution the same way as was presented in reference ~\cite{James:2010vm}. The circular track has a radius of $L = \SI{100}{\metre}$ and lies within the $x-y$ plane. The electron has a velocity of $\beta = 0.999$, corresponding to an energy of \SI{\approx 11.4}{\MeV}. The magnetic field is aligned along the z-axis, with the above parameters resulting in a strength of $\approx 0.3809$\,mT. We place the observer in the same plane as the circle, at a distance of $R = \SI{30}{\km}$ from the center of the circular track to ensure that the calculation takes place in the far-field regime.

We perform this validation in two ways. The first way isolates the test from C8 and its tracking algorithm in order to confirm that the radio module produces correct physical results in stand-alone mode. The second way takes full advantage of C8's functionality, including its tracking algorithm, to ensure that our module generates identical results as an integral part of C8.

For the first case we devised a simple algorithm, which we refer to as ``manual tracking'' that connects \num{100,000} points on a strictly circular track with straight track segments. The second case relies fully on C8's ``leapfrog'' algorithm for particle tracking within the environmental magnetic field, gyrating the electron on a circular track and calculating the synchrotron radiation using C8 as a whole. Using the ``C8 tracking'' we need to configure the tracking algorithm to perform very fine steps (in this case by introducing the next step after a geomagnetic deflection by $10^{-4}$\,rad) for the accurate calculation of the extremely broad-band pulse. In both cases, we allow the electron to perform exactly one revolution.

In figure \ref{fig:synchall}, on the left the pulse for the $y$-polarization produced by the ``manual tracking'' is shown, and on the right the analogous pulse produced by the ``C8 tracking'' is presented. In both test cases the resulting pulses from both the ``CoREAS'' and ``ZHS'' formalisms agree with the analytical solution (detailed in reference \cite{James:2010vm}) to within $\approx 1$\%. The remaining difference at the peak position is related to the finite time binning of the C8 solution. The ``spiky'' contributions visible for the CoREAS formalism are related to interferences between the particle track stepping and the time sampling and will average out for realistic simulations. We thus conclude that both formalisms and the ``C8 tracking'' (for very fine step size) deliver correct results. Also, the observer functionality was implicitly tested this way.

\section{Air Shower comparison}
\label{sec:air-shower-comparison}

\begin{figure}
    \includegraphics[width=8cm]{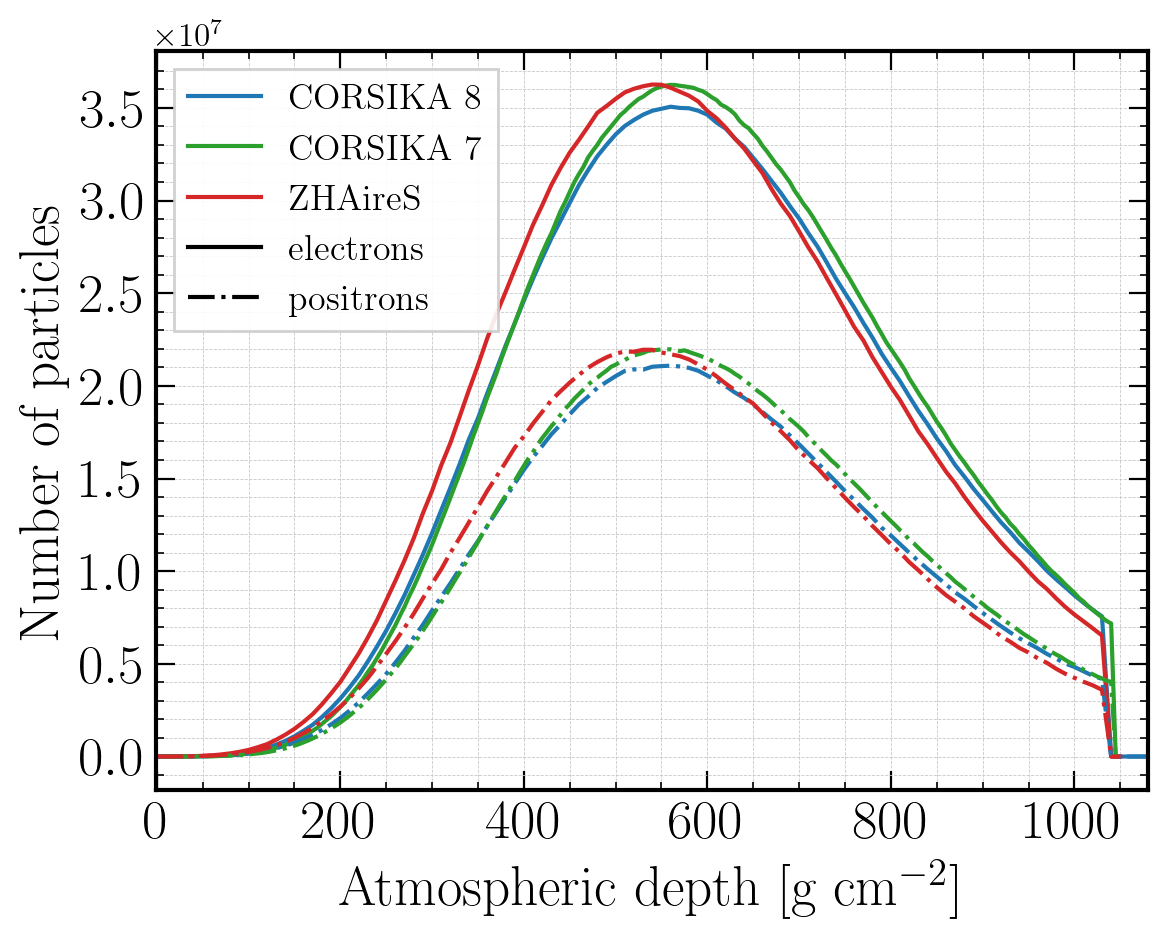}
    \caption{Longitudinal profiles of air showers simulated with C8, C7 and ZHAireS. The number of electrons and positrons as a function of atmospheric depth is shown.}\label{fig:profile}
\end{figure}

\begin{figure*}
    \centering
    \includegraphics[width=0.99\linewidth, trim={3.7cm 4.45cm 3.2cm 4cm}, clip]{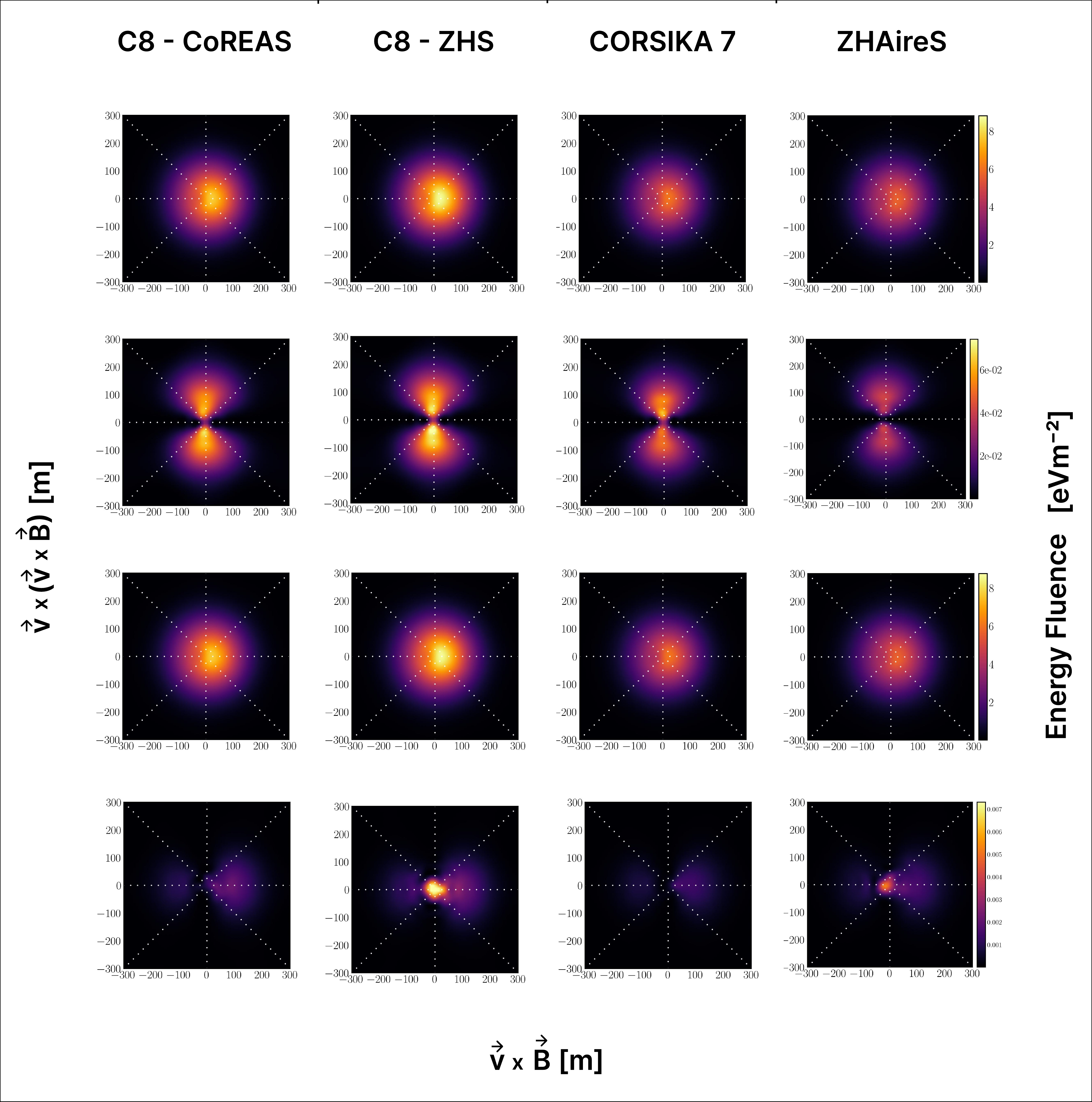}

    \begin{picture}(0,0)
      \put(-210, 510){\small C8 (CoREAS)}
      \put(-80, 510){\small C8 (ZHS)}
      \put(55, 510){\small C7}
      \put(160, 510){\small ZHAireS}
        
      \put(-20, 10){\small $\vec{v} \times \vec{B}$ [m]}
        
      \put(-270, 250){\rotatebox{90}{\small $\vec{v} \times (\vec{v} \times \vec{B})$ [m]}}
        
      \put(250, 235){\rotatebox{90}{\small Energy Fluence [eVm$^{-2}$]}}
    \end{picture}
    
    \caption{Energy fluence maps for different signal polarizations of the electric field in the \SIrange{30}{80}{\mega\hertz} frequency band for C8 with ``CoREAS'' formalism, C8 with ``ZHS'' formalism, C7 with CoREAS extension and ZHAireS. The order of the polarizations we see starting from top to bottom is: absolute value, $ \vec{v} \times (\vec{v} \times \vec{B}) $, $\vec{v} \times \vec{B}$ and $\vec{v}$, where $\vec{v}$ corresponds to the shower axis and $\vec{B}$ corresponds to the magnetic field axis. Please note the strongly different z-axis scales.}
    \label{fig:maps}
\end{figure*}

Following the successful validation of the radio module, we proceed with simulating the radio emission from a vertical air shower and compare our results with C7 and ZHAireS. For the shower simulation we choose the primary particle to be an iron nucleus with a primary energy of \SI{100}{\peta\eV}. We use an energy cut of \SI{0.5}{\MeV} for the electromagnetic particles and \SI{0.3}{\giga\eV} for hadrons and muons. The same hadronic interaction models are used for C8, C7 and ZHAireS, namely Sibyll 2.3d~\cite{Riehn:2019jet} for high energies and FLUKA~2021.2.7~\cite{Battistoni:2015epi} for low energy particles. For C8, electromagnetic interactions are handled by PROPOSAL v7.6.2~\cite{koehne2013proposal, dunsch_2018_proposal_improvements, proposal-zenodo}. To reduce computing times we used thinning in all 3 simulation codes: in C7 using a thinning threshold of $10^{-6}$ with a maximum weight of 100, in C8 using a thinning threshold of $10^{-6}$ and a maximum weight of 50, and in ZHAireS using a thinning threshold of $10^{-7}$ and a \textit{ThinningWFactor} of 5, which led to similar runtimes as in C7. We stress that the thinning implementations vary between the codes. The algorithm implemented within C8 is detailed in reference \cite{CORSIKA:2023jyz}; in comparison with the one used in C7 it requires half the maximum weight for equivalent thinning quality. For the environment configuration in all simulations we used Linsley's parameterization of the ``US Standard Atmosphere'', we set the refractive index at sea level to $n=\num{1.000327}$, scaling with density according to the Gladstone-Dale law. The geomagnetic field was set as horizontal with a value of \SI{50}{\micro\tesla}. For the observer collection we use a star-shaped pattern of 160 observers located at sea level, in 20 concentric rings spaced equally from \SIrange{25}{500}{\metre} from the shower axis with 8 observers distributed equidistant in azimuth in each ring. The sampling period is set to \SI{0.1}{\nano\second}, allowing to safely evaluate frequencies up to 1~GHz. It is worth pointing out that for C8 we run one shower simulation which employed two \textit{radio processes}. Both used the same filters, propagator, and identical observer collections and differ only in the formalism, one used the ``CoREAS'' and the other one the ``ZHS'' formalism. Particle tracking was left at standard values in ZHAireS and C7 (STEPFC parameter at 1, maximum geomagnetic deflection before adding another step is 0.2 rad.). For C8, the maximum allowed geomagnetic deflection before adding another tracking step was also set to 0.2 rad.

For this comparison, we first ran several showers with all simulation codes and then picked one simulation per code so that the three of them were similar in terms of longitudinal shower development, see \cref{fig:profile}. They are similar in terms of number of electrons, positrons and depth of shower maximum, $X_{\rm max}$. For C8 $X_{\rm max}$ is $\sim\SI{560}{\gram \per \cm \squared}$, for C7 $\sim\SI{570}{\gram \per \cm \squared}$ and for ZHAireS $\sim\SI{550}{\gram \per \cm \squared}$. We proceed and compare the radio emission from these showers. When comparing the simulations from C8, C7 and ZHAireS one must keep in mind that each shower code uses different tracking and thinning algorithms as well as different electromagnetic interaction models, so it is unrealistic to expect a complete agreement.

Qualitative agreement for all observers between the three codes can be examined best through the energy fluence (energy deposited by radio waves per unit area in a given frequency band) predicted at the ground. For this, we plot 2D fluence maps in the \SIrange{30}{80}{\MHz} frequency band, a common band used by radio-detection experiments, cf. \cref{fig:maps}, in the shower plane (equal to ground plane for vertical incidence). The exact positions of observers are indicated with white dots at which the fluence is directly calculated, while for positions in between fluence values have been interpolated. The emission footprints predicted by C8 using both formalisms, C7, and ZHAireS agree well in terms of shape and symmetry of the energy fluence. It is evident, though, that the C8 shower simulation for both formalisms predicts an energy fluence roughly 25\% higher than those predicted by C7 and ZHAireS. As we do not compare identical showers, no perfect agreement is expected. The vertical polarization in the 2D fluence maps shows an interesting feature. We can see that the ``ZHS'' formalism, both in C8 and ZHAireS predicts a ``blip'' near the shower axis, which is absent in the calculation with the ``CoREAS'' formalism. In practical terms, though, polarization along the shower axis is not relevant since the fluence in it is negligible.

We have studied the lateral profiles of electrons and positrons in air showers as simulated by C8 and C7 and have found that showers simulated with C8 for these standard settings generally exhibit more particles close to the shower axis than those simulated with C7 \cite{AlameddinePhD}. This could explain why we see more fluence from C8 showers, even when the longitudinal profiles of the showers are very similar. More particles near the shower core would make the radiation emitted more coherent and hence produce the higher levels of fluence that we observe in the Cherenkov ring and within it. 

In the following, we proceed to investigate details of the simulation configuration in order to work out the achievable level of agreement between the codes for optimized simulation settings.

\section{Detailed formalism comparison for identical showers}
\label{sec:coreas-vs-zhs}

\begin{figure*}
\centering
\begin{subfigure}{.5\textwidth}
    \centering
    \includegraphics[width=7.5cm]{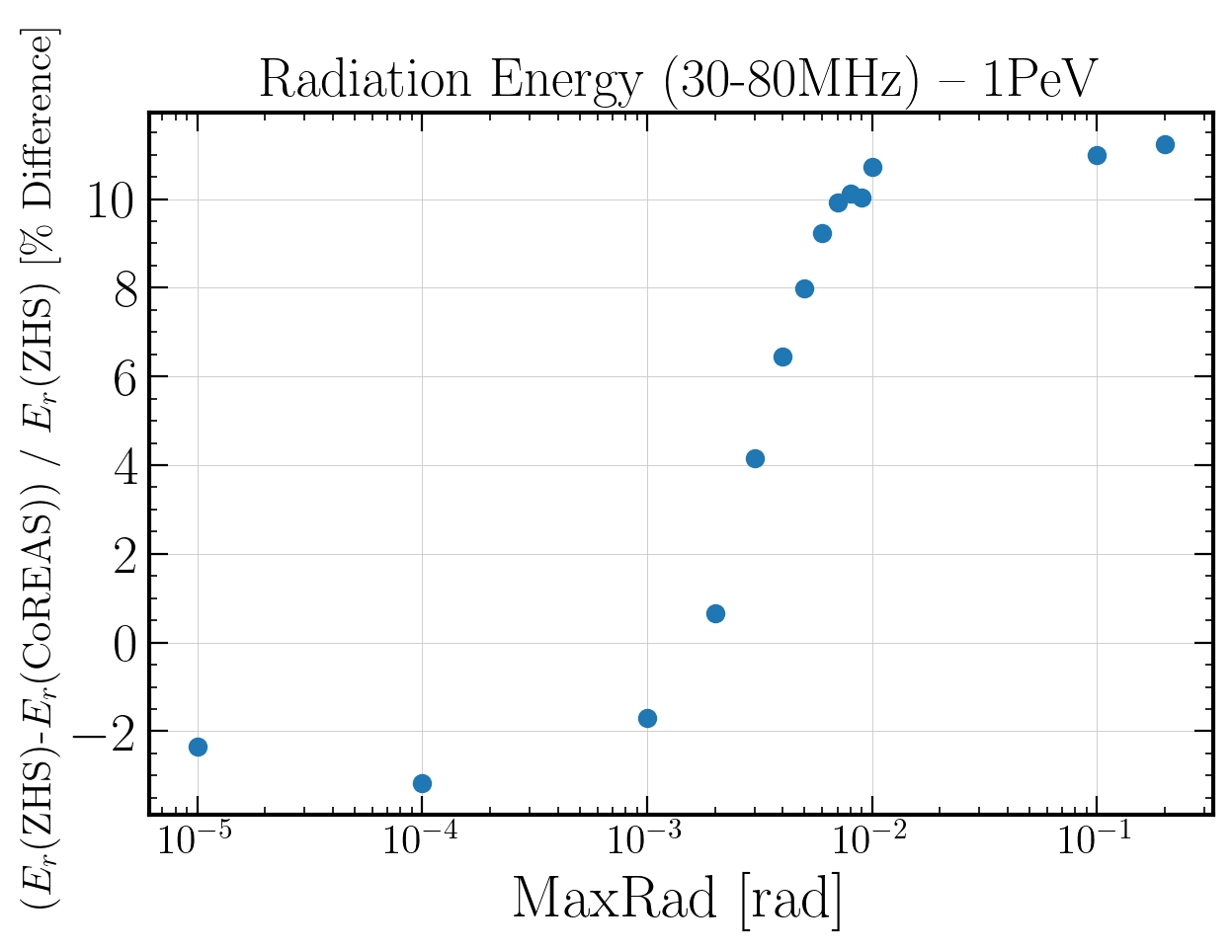}
\end{subfigure}%
\begin{subfigure}{.5\textwidth}
    \centering
    \includegraphics[width=7.5cm]{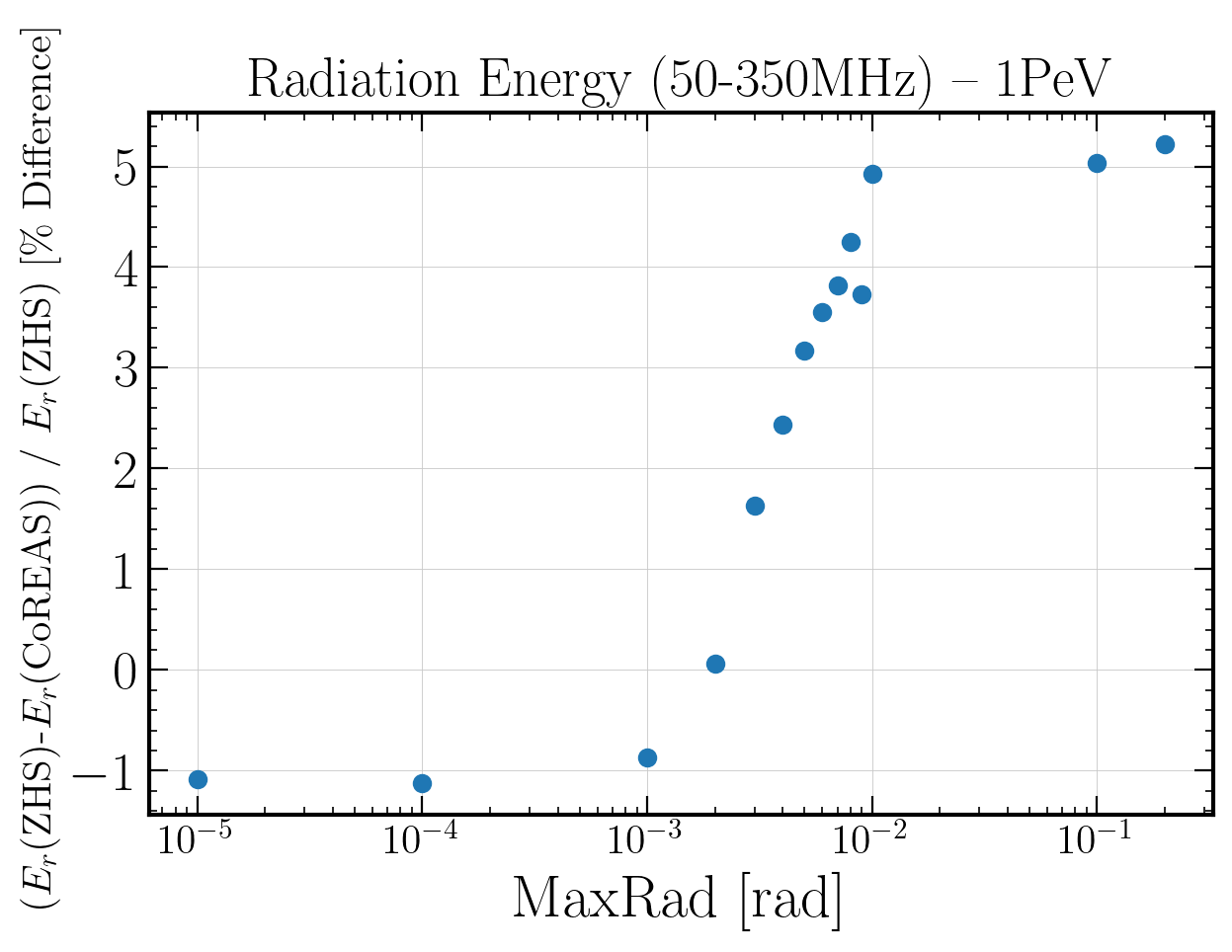}
\end{subfigure}
\caption{Level of agreement between the radiation energy predicted by C8 simulations with the ``CoREAS'' and ``ZHS'' formalisms as a function of the maximally allowed magnetic deflection angle (MaxRad), which effectively alters the simulated particle track length. The comparison between the radiation energy predicted by the two formalisms is evaluated in the \SIrange{30}{80}{\mega\hertz} (left) and \SIrange{50}{350}{\mega\hertz} (right) frequency bands.}
\label{fig:coreas-zhs-maxRad-radiationenergy}
\end{figure*}

\begin{figure*}
    \centering
    ($f$(ZHS) - $f$(CoREAS)) / $f$(ZHS),  \SIrange{30}{80}{\mega\hertz}\\

    \vspace{1em} 

    \threeByTwoGrid
    {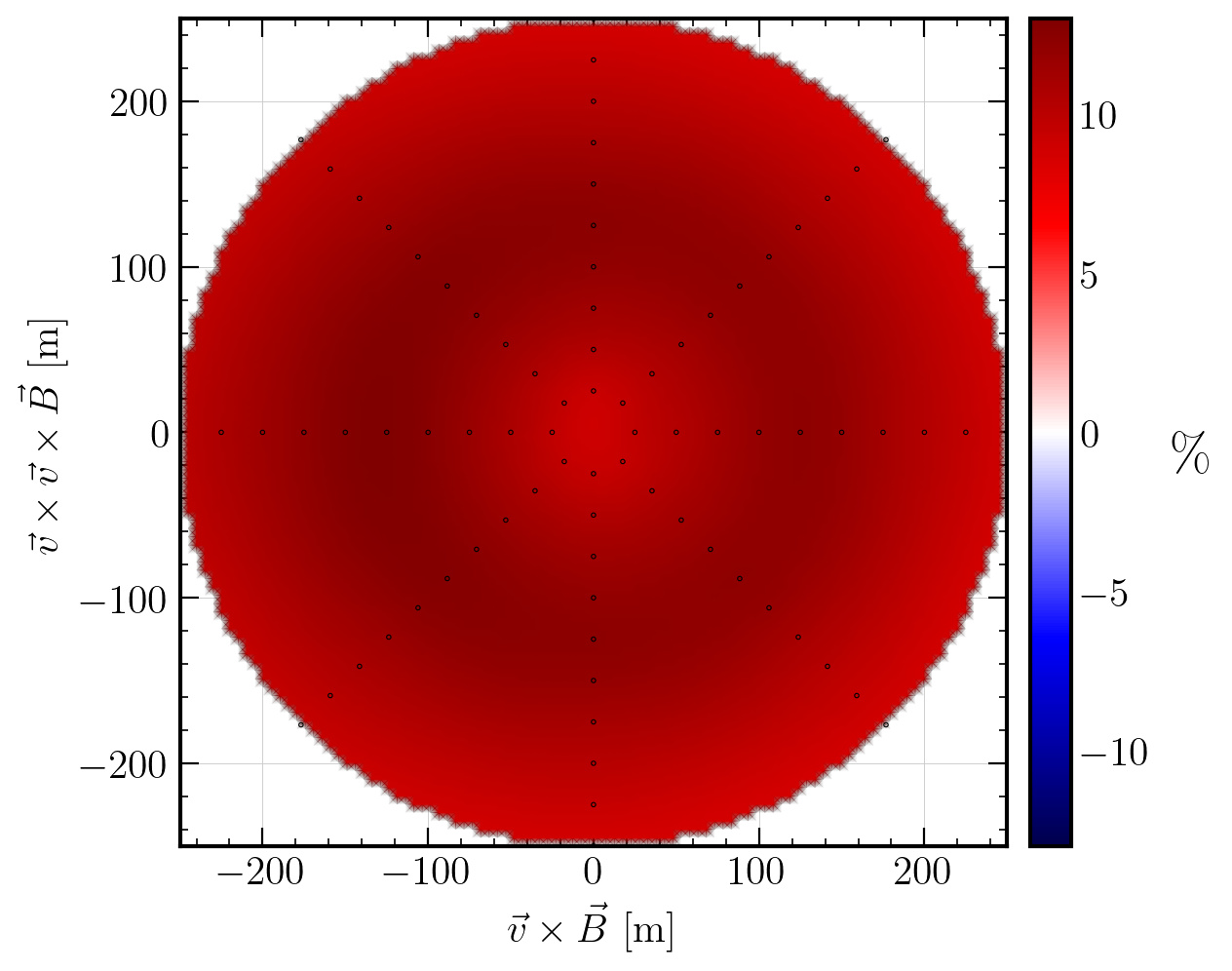}
    {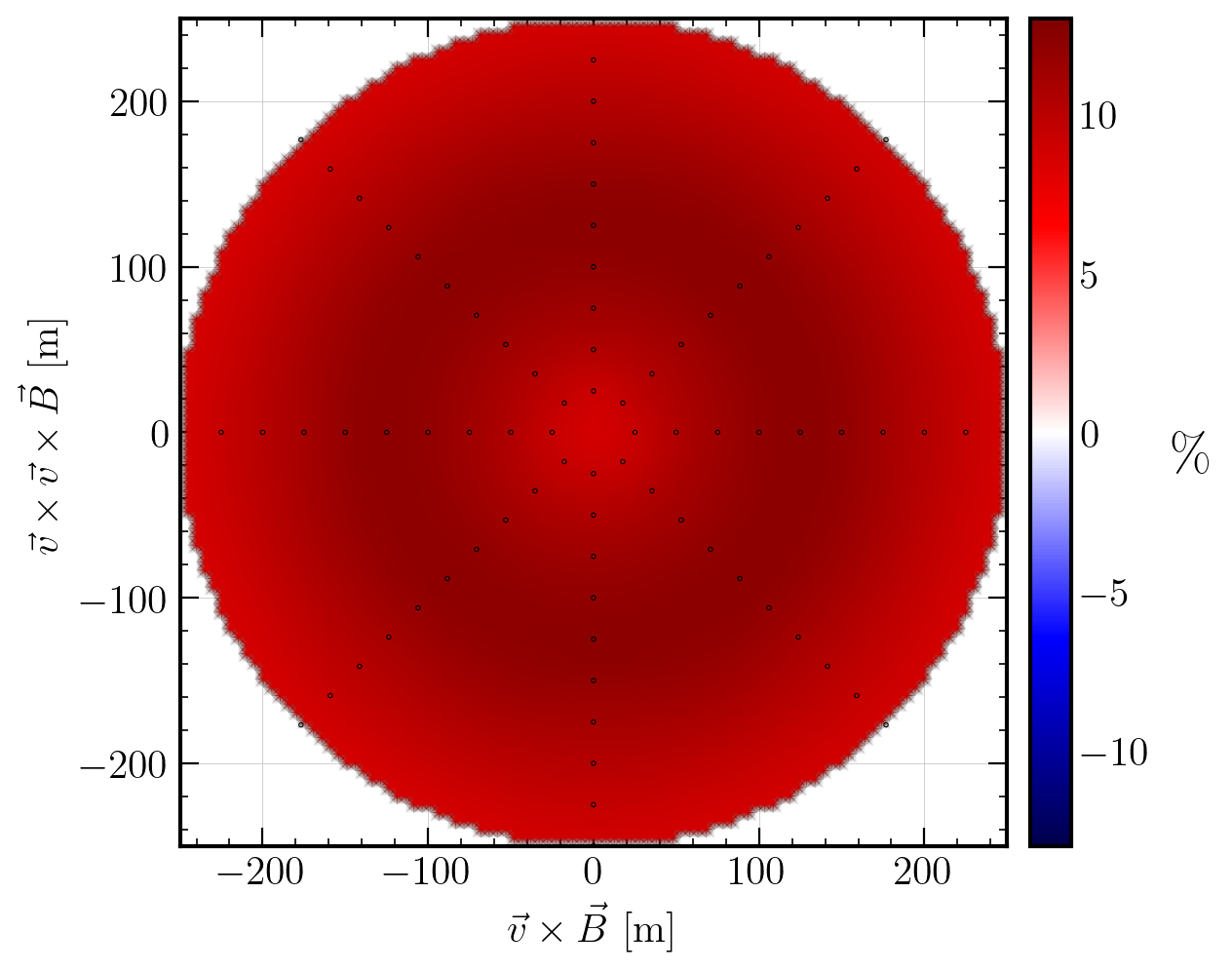}
    {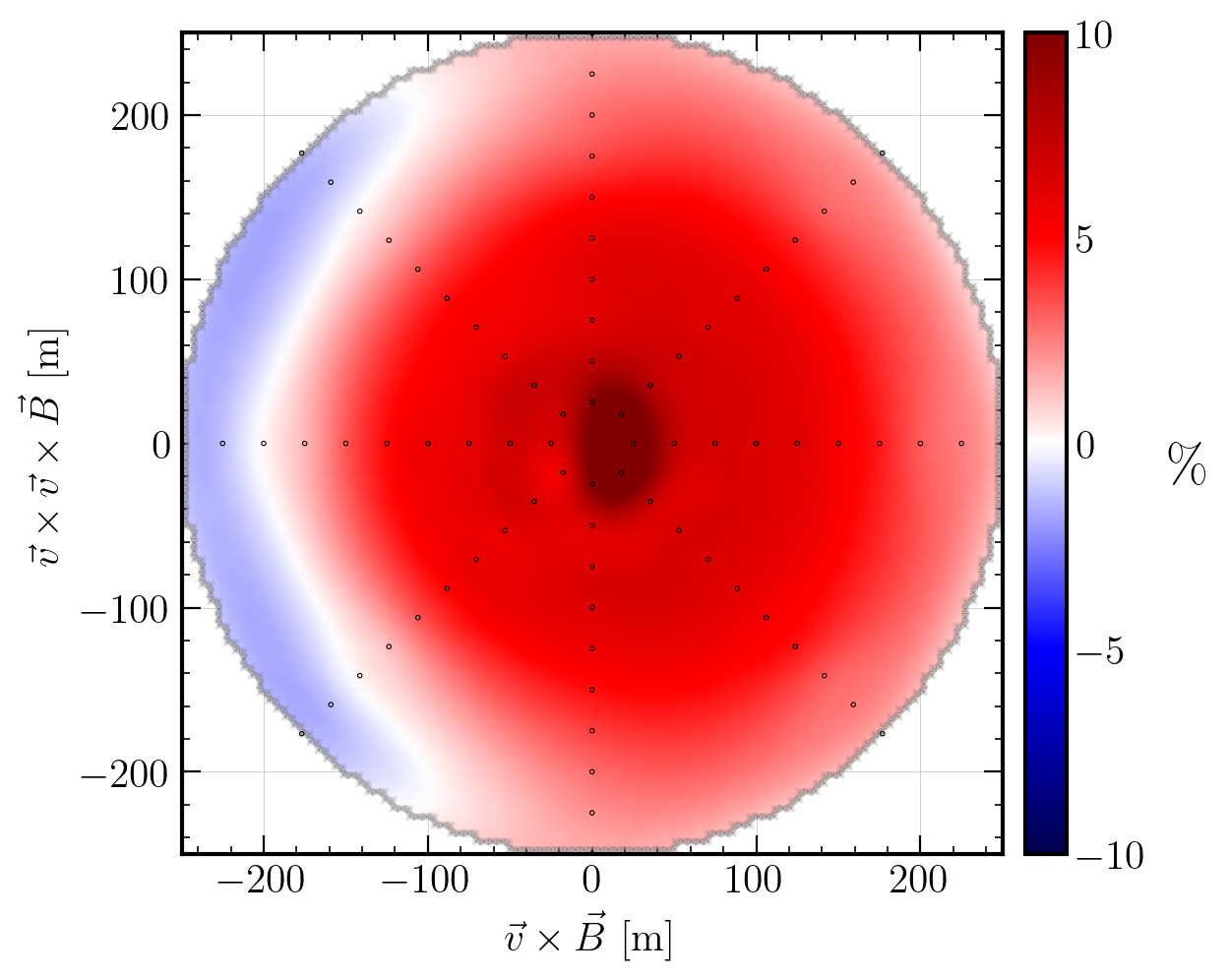}
    {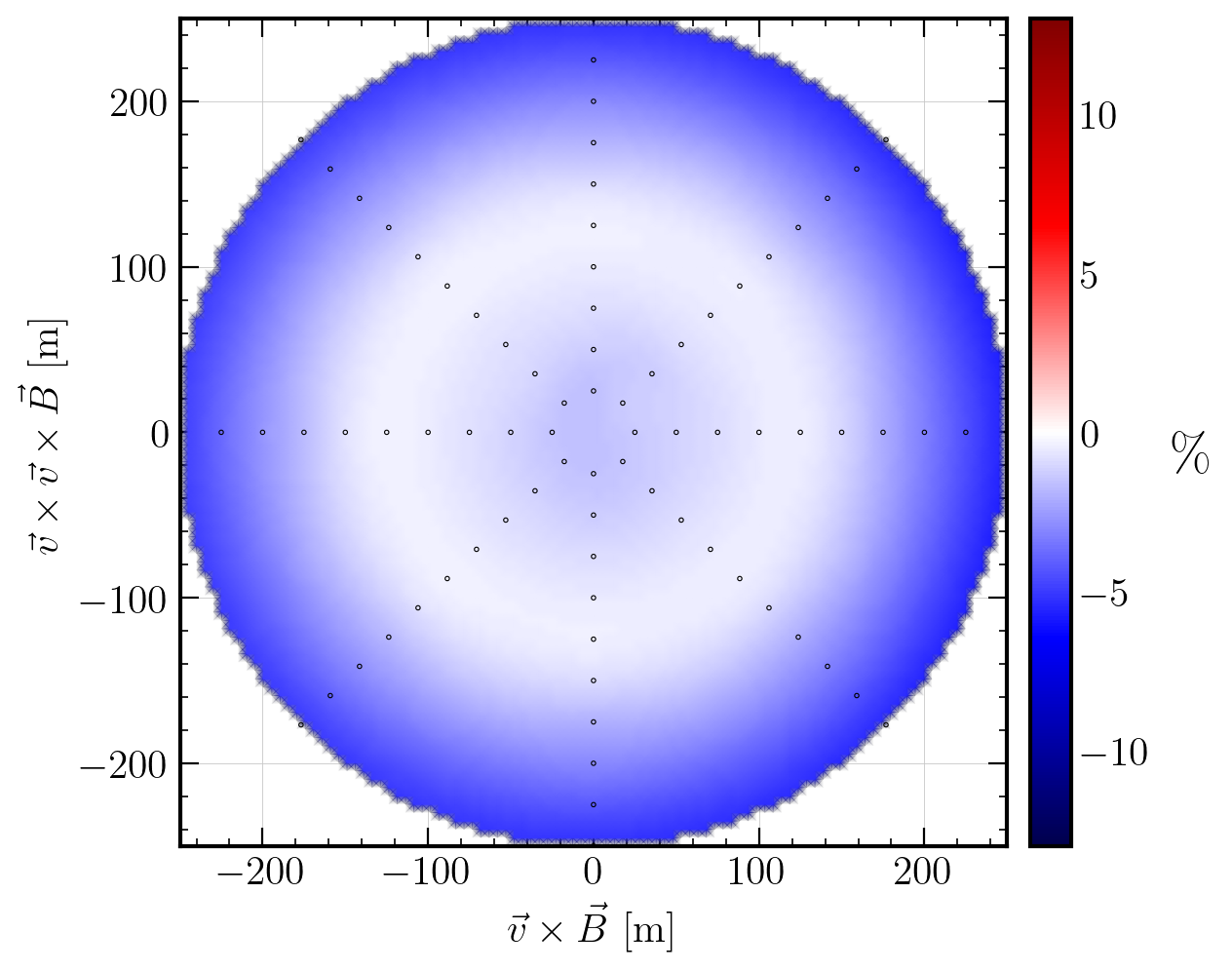}
    {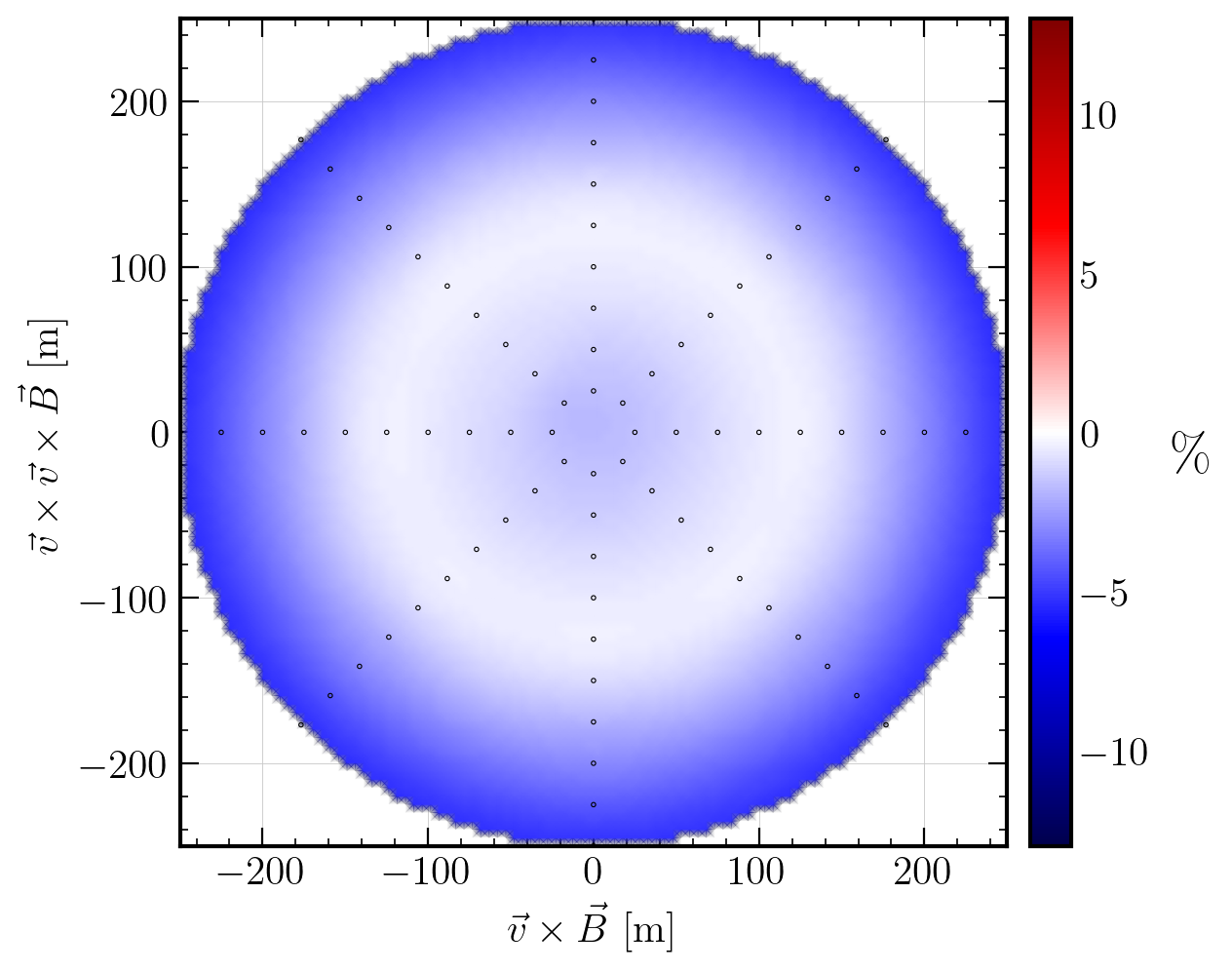}
    {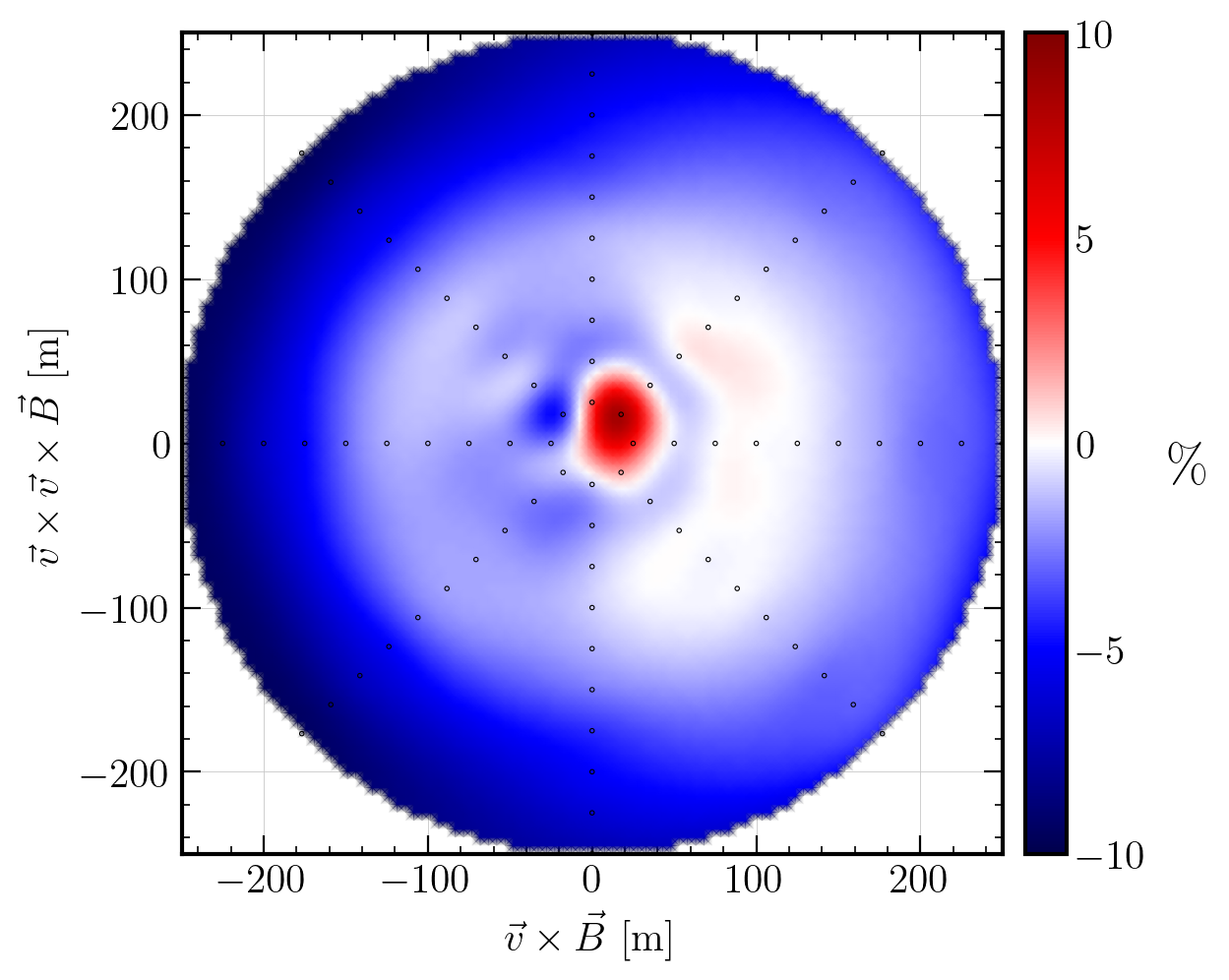}
    
\caption{Percentage deviation maps for the energy fluence simulated with the ``ZHS'' and ``CoREAS'' formalisms averaged over 100 iron-induced vertical air showers with an energy of \SI{1}{\peta\eV} in the \SIrange{30}{80}{\mega\hertz} frequency band. Upper row is for MaxRad value of 0.2, lower row is for MaxRad value of 0.001. From left to right: total fluence, geomagnetic fluence, and charge-excess fluence.}
\label{fig:coreas-vs-zhs-ratiomaps-30-80}
\end{figure*}

\begin{figure*}
    \centering
    ($f$(ZHS) - $f$(CoREAS)) / $f$(ZHS),  \SIrange{50}{350}{\mega\hertz}\\

    \vspace{1em} 

    \threeByTwoGrid
    {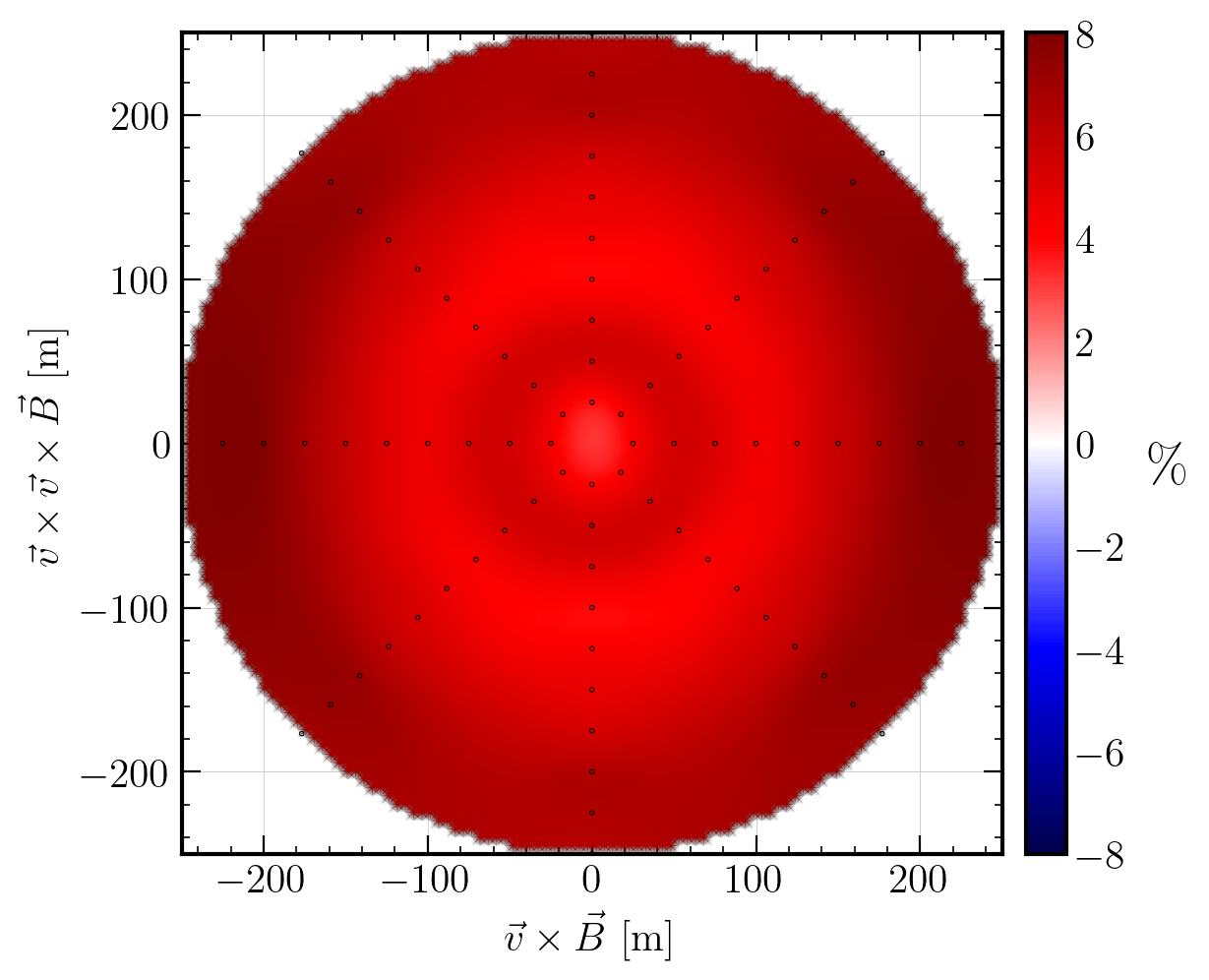}
    {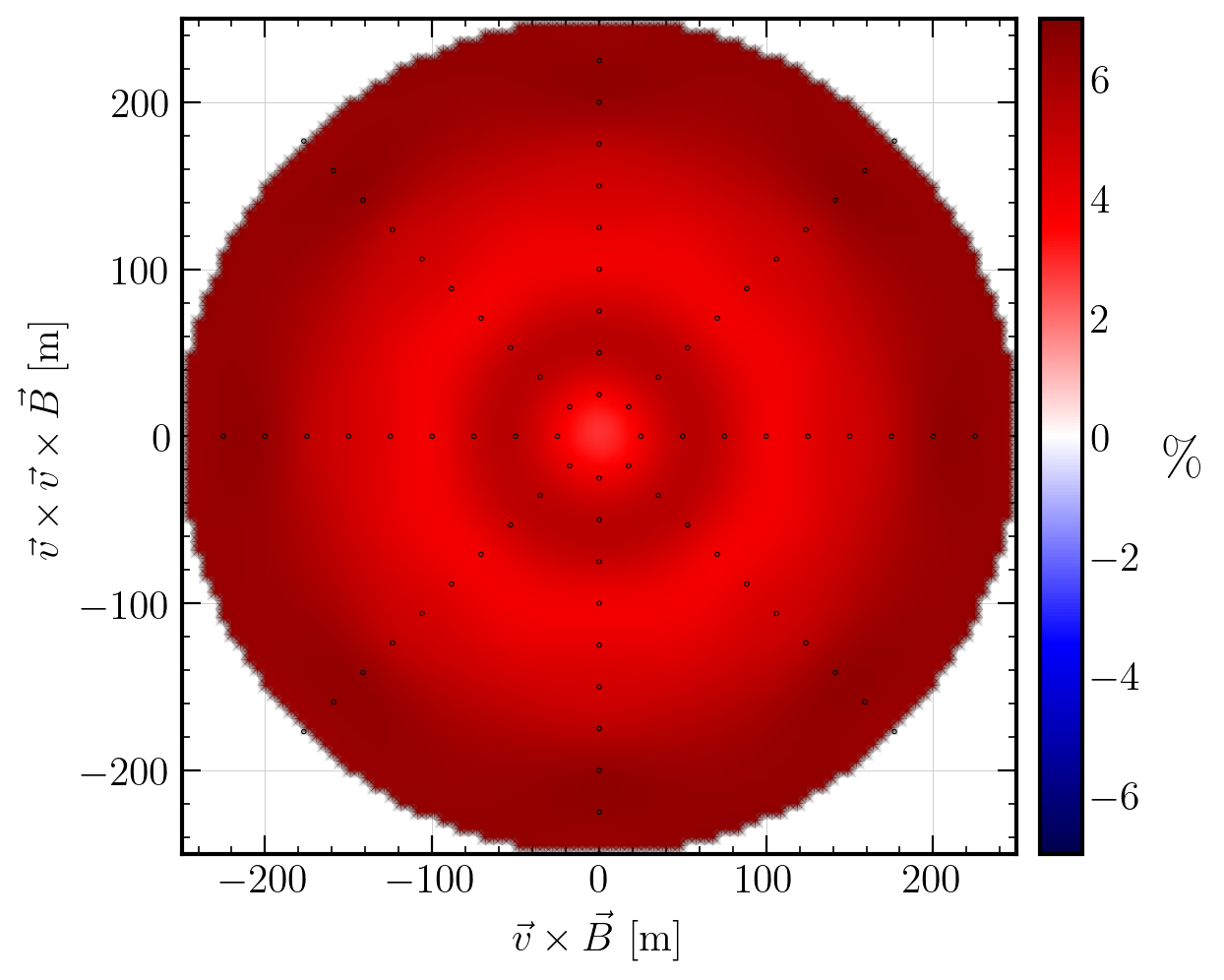}
    {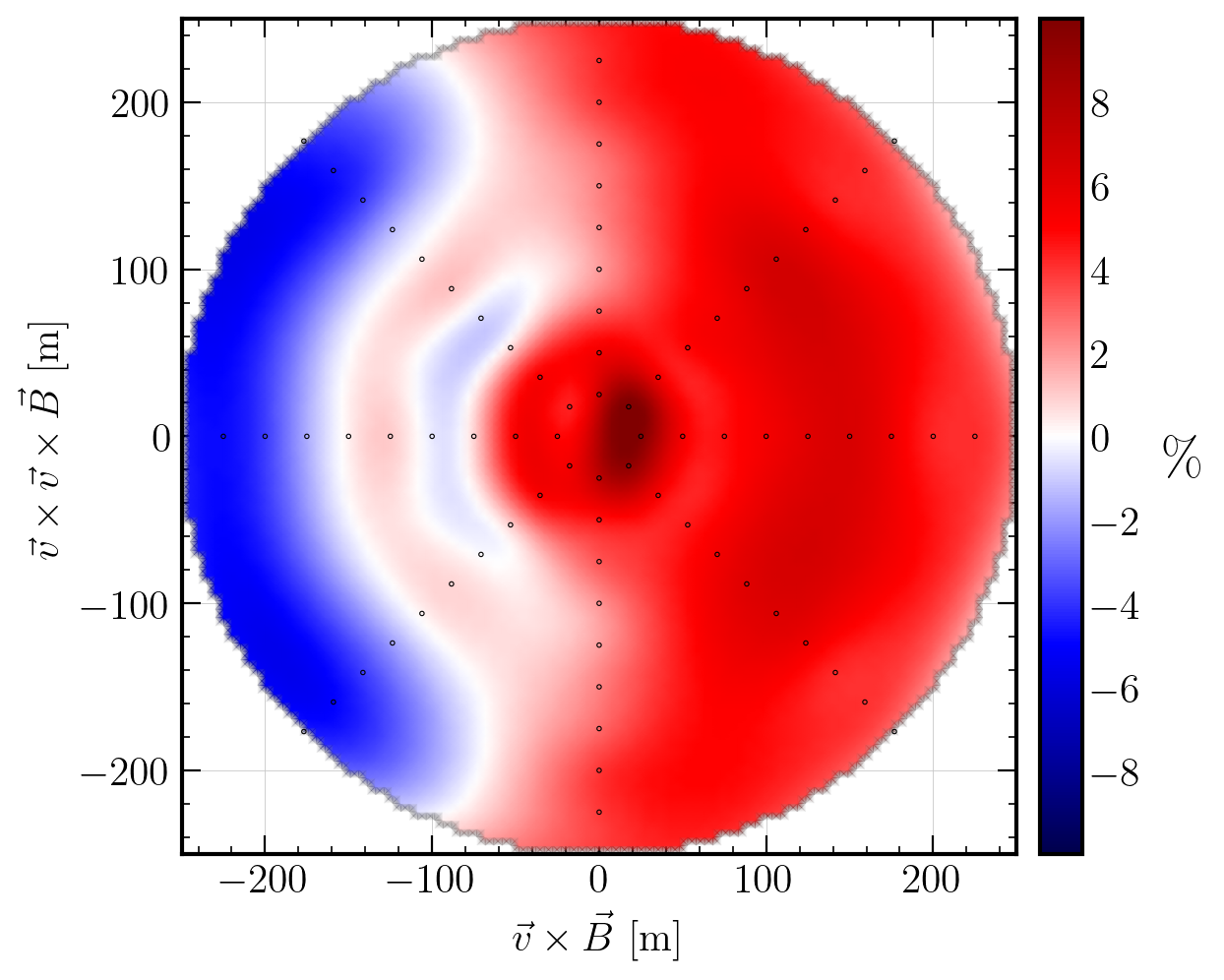}
    {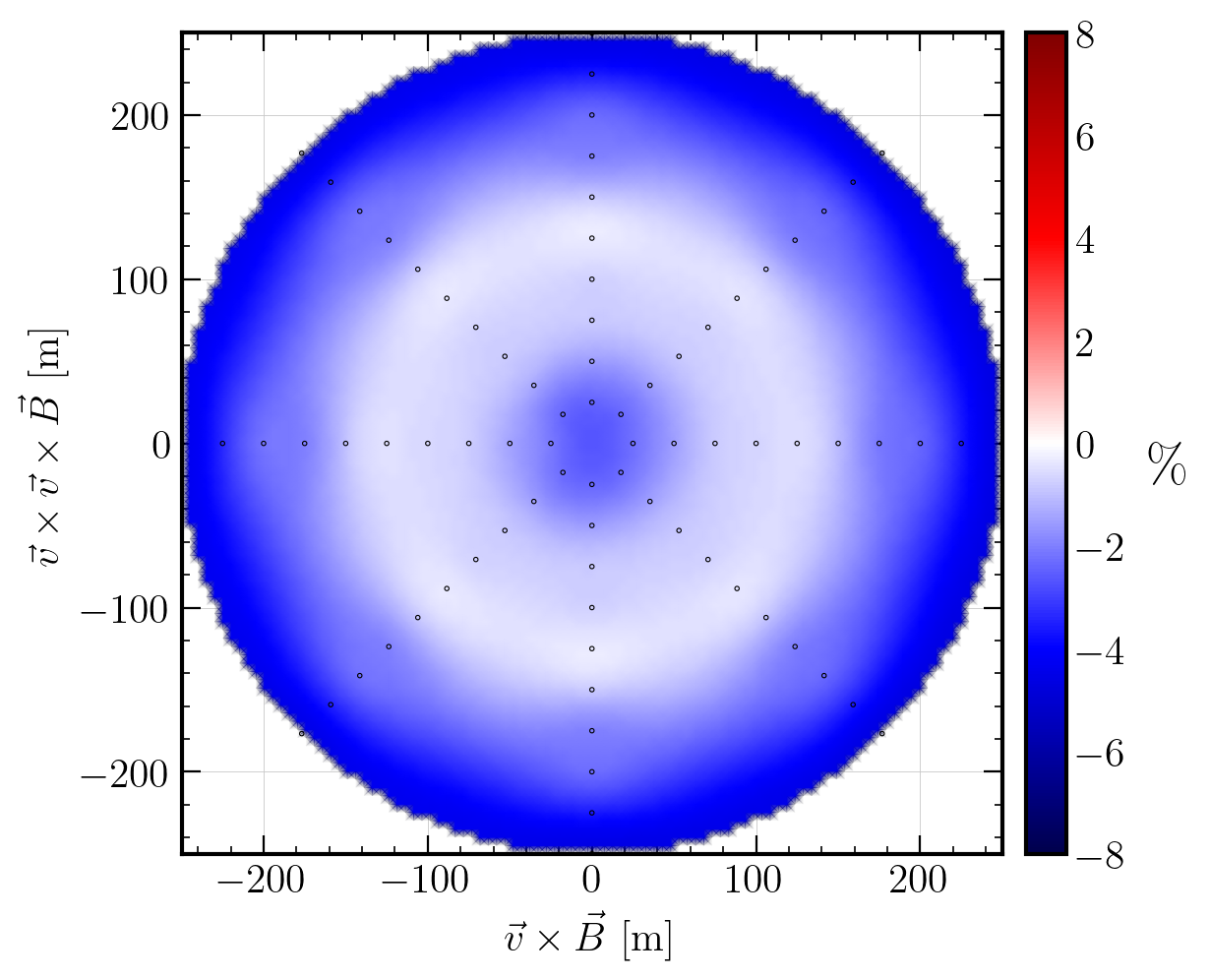}
    {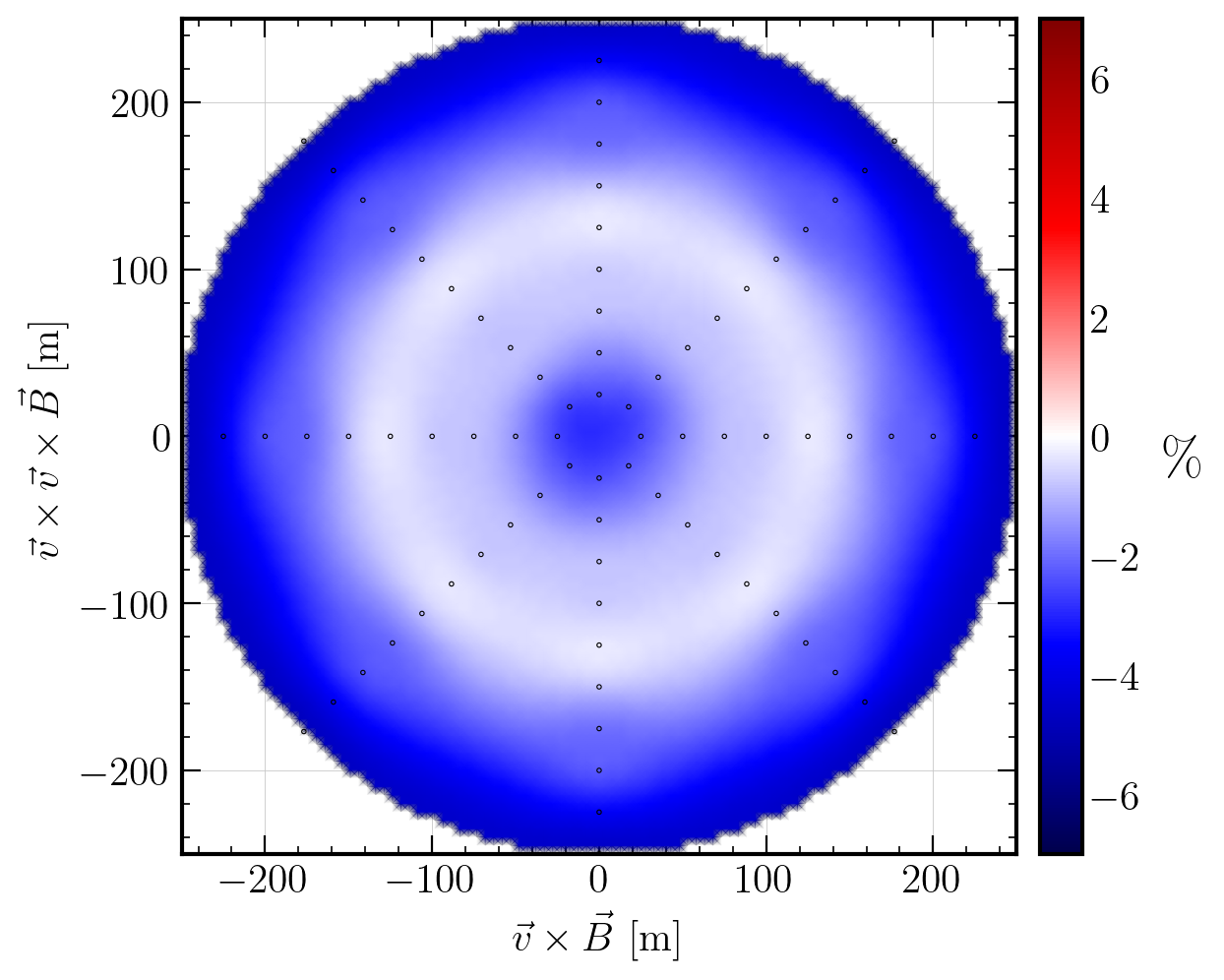}
    {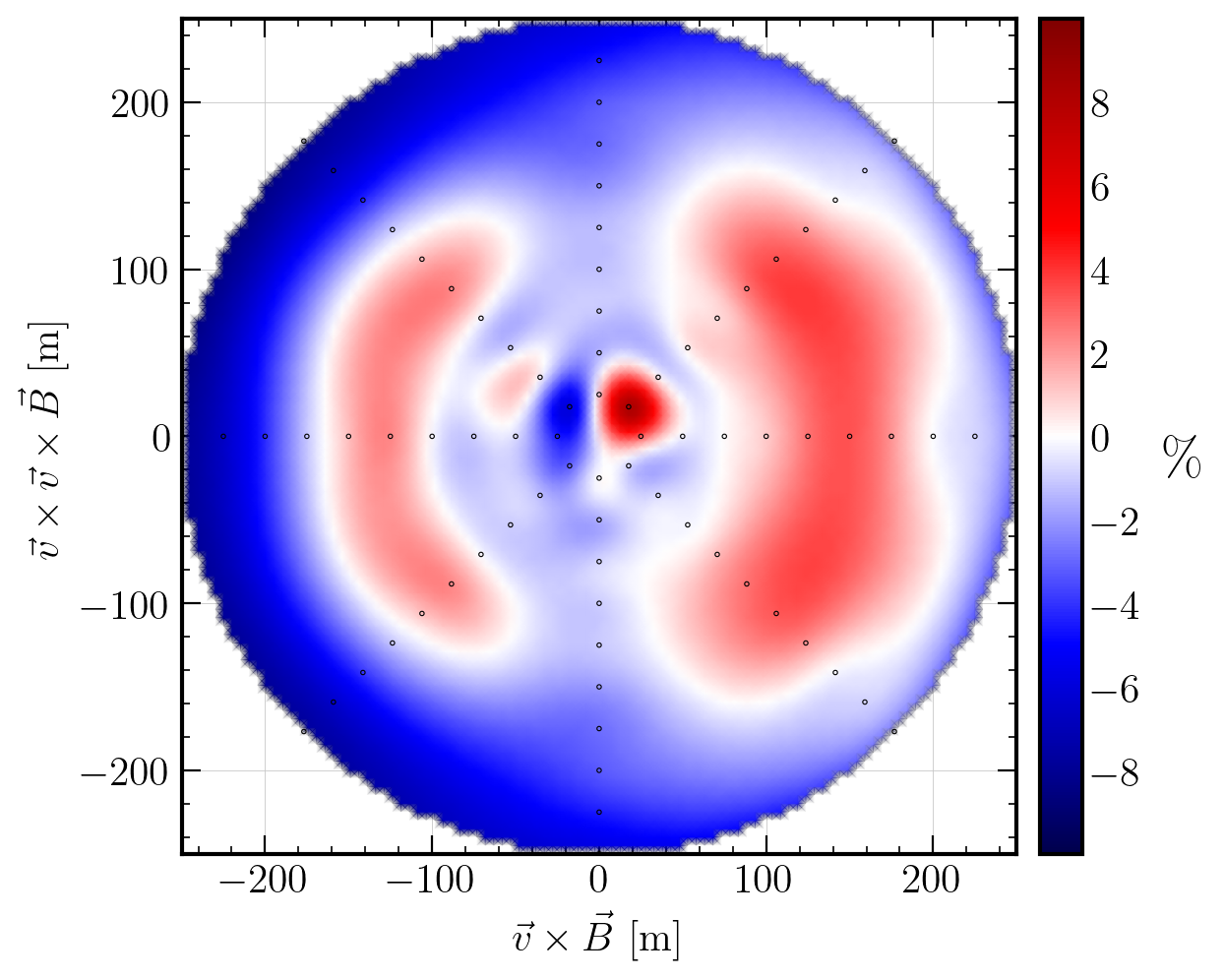}
    
\caption{Same as figure \ref{fig:coreas-vs-zhs-ratiomaps-30-80} but for the 50-350~MHz band.}
\label{fig:coreas-vs-zhs-ratiomaps-50-350}
\end{figure*}

\begin{table}
\centering
\begin{tabular}{l
                S[table-format=+2.1, table-space-text-post={\,\%}] 
                S[table-format=+2.1, table-space-text-post={\,\%}]}
\toprule
\textbf{Frequency Band} & \multicolumn{2}{c}{\textbf{MaxRad Value}} \\
\cmidrule(lr){2-3}
& \textbf{0.2 rad} & \textbf{0.001 rad} \\
\midrule
\textbf{30-80\,MHz} &   &  \\
Total Emission & 11.3\,\% & 1.4\,\% \\
Geomagnetic Emission & 11.2\,\% & -1.5\,\% \\
Charge Excess Emission & 4.0\,\% & -2.8\,\% \\
\midrule
\textbf{50-350\,MHz} &   &   \\
Total Emission & 4.9\,\% & -1.0\,\% \\
Geomagnetic Emission & 4.6\,\% & -1.0\,\% \\
Charge Excess Emission & 3.1\,\% & 0.6\,\% \\
\bottomrule
\end{tabular}
\caption{Percentage difference in radiation energy between C8 simulations with the ``CoREAS'' and ``ZHS'' formalisms for different frequency bands and MaxRad settings. Positive values indicate more radiation energy from the ``ZHS'' formalism.}
\label{tbl:coreas-vs-zhs-fluence}
\end{table}

Multiple \textit{Radio Processes} can be used in a C8 shower simulation. This gives the user the capability to perform the radio-emission calculations for the exact same shower with more than one formalism. In this section, the ``CoREAS'' and ``ZHS'' formalisms are directly compared. Good agreement between the two formalisms would underline that radio emission from particle showers is well-understood and that the simulation predictions can form a solid foundation for the development of analysis approaches for the radio detection technique.

One particular question of interest is how the agreement between the two formalisms depends on details of the setup of the simulation. Here, we quantify the level of agreement between the two formalisms in terms of the predicted radiation energy\footnote{The area integral over the energy fluence footprint, i.e., the total energy deposited on the ground by radio waves in a given frequency band.} \cite{Glaser:2016qso} as a function of the maximally allowed magnetic deflection angle (MaxRad) in the simulation. MaxRad is a parameter within the tracking algorithm of C8 (and C7) governing the maximum length for a curved track approximated by a straight line. We run 14 shower simulations with essentially the same setup as described in the previous section. However, we lower the particle energy to 1\,PeV, do not apply thinning, and reduce the number of observers in the star-shape pattern to 80 observers located on 10 concentric rings spaced equally from \SIrange{25}{250}{\metre} from the shower core. The individual showers only differ in terms of their MaxRad value and their longitudinal profiles (as the random number sequence changes with the MaxRad setting).

It is shown in \cref{fig:coreas-zhs-maxRad-radiationenergy}, that as the particle track length gets smaller (smaller values of MaxRad), the agreement between the two formalisms in terms of predicted radiation energy improves until it reaches a plateau for MaxRad values smaller than \num{0.001} rad. Here, the results from the ``CoREAS'' and ``ZHS'' formalisms agree within $2\%$ in the \SIrange{30}{80}{\mega\hertz} and within $1\%$ in the \SIrange{50}{350}{\mega\hertz}, a higher-frequency range targeted in particular by the Square Kilometre Array. We note that the results for \textit{both} formalisms change when the MaxRad value is decreased; the value for the ``ZHS'' formalism changes more substantially, until the results from both formalisms converge. (We note that it is not straight-forward to compare absolute radiation-energy values between different MaxRad settings, as the random number sequence changes and thus the showers are not directly comparable.)

We note furthermore that with the radio module in C8, one can study the level of agreement of the two formalisms also in other media, in particular ice. It is worth noting in this context that in denser media than air the ``ZHS-like'' fallback within the ``CoREAS'' formalism needs to be adjusted to accomodate the much larger Cherenkov angle, see also \cite{Bechtol:2021tyd}.

Of the MaxRad values simulated here, we investigate two relevant cases further, namely the case of 0.2 rad which is the default value for the tracking algorithm and the 0.001 rad where the agreement reaches an equilibrium. We produce 100 simulations with the same settings as described in the previous paragraph for each of the two MaxRad values. Using these 100 simulations, we compare the results predicted by the two formalisms in terms of their fluence footprints and their predicted radiation energy.

To calculate and plot the mean total fluence per observer $i$, $\overline{f_{\text{total}}^i}$, for the ``COREAS'' and ``ZHS'' formalism for all 100 showers in each MaxRad case, the following formula is used:

\begin{equation}
    \overline{f_{\text{total}}^i} = \epsilon_0 c \Delta t \sum_{j=1}^{100} \sum_{t_{\mathrm{start}}}^{t_{\mathrm{end}}}E_{ji}(t)^{2} / 100,
\label{eq:many-fluence}
\end{equation}
where $j$ represents the shower number. Using the same notation, the following formulas are used to calculate and plot the mean values of the geomagnetic contribution, $\overline{f_{\text{geo}}^i}$, and the charge-excess contribution, $\overline{f_{\text{ce}}^i}$, in shower plane coordinates \cite{Glaser:2016qso}:

\begin{equation}
    \overline{f_{\mathrm{geo}}^i} = \sum_{j=1}^{100} \biggl( \sqrt{f^{ji}_{\Vec{v} \times \Vec{B}}} - \frac{\cos\phi}{|\sin\phi|} \cdot \sqrt{f^{ji}_{\Vec{v} \times (\Vec{v} \times \Vec{B})}}  \biggr)^2 / 100
\label{eq:many-fluence-geo}
\end{equation}

\begin{equation}
    \overline{f_{\mathrm{ce}}^i} = \sum_{j=1}^{100} \biggl( \frac{1}{\sin^2\phi} \cdot f^{ji}_{\Vec{v} \times (\Vec{v} \times \Vec{B})} \biggr) / 100.
\label{eq:many-fluence-ce}
\end{equation}
Here, $\phi$ is the polar angle between the observer position and the positive $\Vec{v} \times \Vec{B}$ axis. For $\phi = 0$ or $\phi = \pi$ the $1/\sin\phi$ term diverges. To tackle this, an interpolation takes place where these values are calculated as the mean value from their nearest neighbors on the same radial distance from the shower core.

Ratio maps for all showers in terms of percentage are calculated over the mean values of fluence for the ``CoREAS'' and ``ZHS'' formalisms. The percentage difference per observer $i$, $D_i$, is calculated using the ``ZHS'' formalism as reference. This is calculated for the total signal, the geomagnetic contribution, and the charge-excess emission. If the difference turns out positive that indicates that the ``ZHS'' formalism produces higher fluence than the ``CoREAS'' formalism (red color in \cref{fig:coreas-vs-zhs-ratiomaps-30-80} and \cref{fig:coreas-vs-zhs-ratiomaps-50-350}).

The ratio maps that showcase the level of agreement of the two formalisms for total fluence, geomagnetic contribution, and charge-excess emission in the \SIrange{30}{80}{\mega\hertz} and \SIrange{50}{350}{\mega\hertz} frequency bands are shown in \cref{fig:coreas-vs-zhs-ratiomaps-30-80} and \cref{fig:coreas-vs-zhs-ratiomaps-50-350}, respectively. For the \SIrange{30}{80}{\mega\hertz} band the deviations are larger for longer particle tracks, i.e. MaxRad of 0.2 rad. The ``ZHS'' formalism predicts higher fluence than the ``CoREAS'' formalism of up to $12\%$ in regions where the signal is strong, i.e. close to the Cherenkov ring, in both total fluence and geomagnetic contribution. The agreement for the charge-excess footprint is different with even a region for which the ``CoREAS'' formalism predicts higher fluence than the ``ZHS'' formalism (blue-colored region) by less than $3\%$. The highest differences are observed close to the shower core where the ``ZHS'' formalism predicts up to $10\%$ higher fluence.

In the case of small particles tracks (MaxRad = 0.001 rad), cf. \cref{fig:coreas-vs-zhs-ratiomaps-30-80} bottom row, the agreement is much better. In regions where the signals are strong, the two formalisms agree within $2\%$. Regarding the charge excess fluence footprint the trend is similar, with the most interesting point being that the ``ZHS'' formalism predicts up to $9\%$ higher fluence at a certain position very close to the shower core.

In the \SIrange{50}{350}{\mega\hertz} frequency band (\cref{fig:coreas-vs-zhs-ratiomaps-50-350}), the overall agreement of both formalisms is better for both MaxRad values than in the \SIrange{30}{80}{\mega\hertz} band. For longer particle tracks, in the total fluence and geomagnetic contribution fluence ratio maps, the deviations reach a maximum of roughly $6\%$ at larger distances. Around the Cherenkov ring, where the signal is strong, the agreement is within $~4\%$. For the shorter particle tracks, the deviations reach a maximum of $~3\%$ at larger distances. In the regions close to the Cherenkov ring, the two formalisms agree within $1\%$.

An area integration over the fluence maps yields the radiation energy, for which the level of agreement between the two formalisms is summarized in \cref{tbl:coreas-vs-zhs-fluence}. These values confirm once more that for very small particle tracks the two formalisms practically converge also for the set of 100 showers.

\section{Comparison between C8 and C7 for small tracks}
\label{sec:c8-vs-c7}

\begin{figure*}
\centering
\begin{subfigure}{.5\textwidth}
    \centering
    \includegraphics[width=7.5cm]{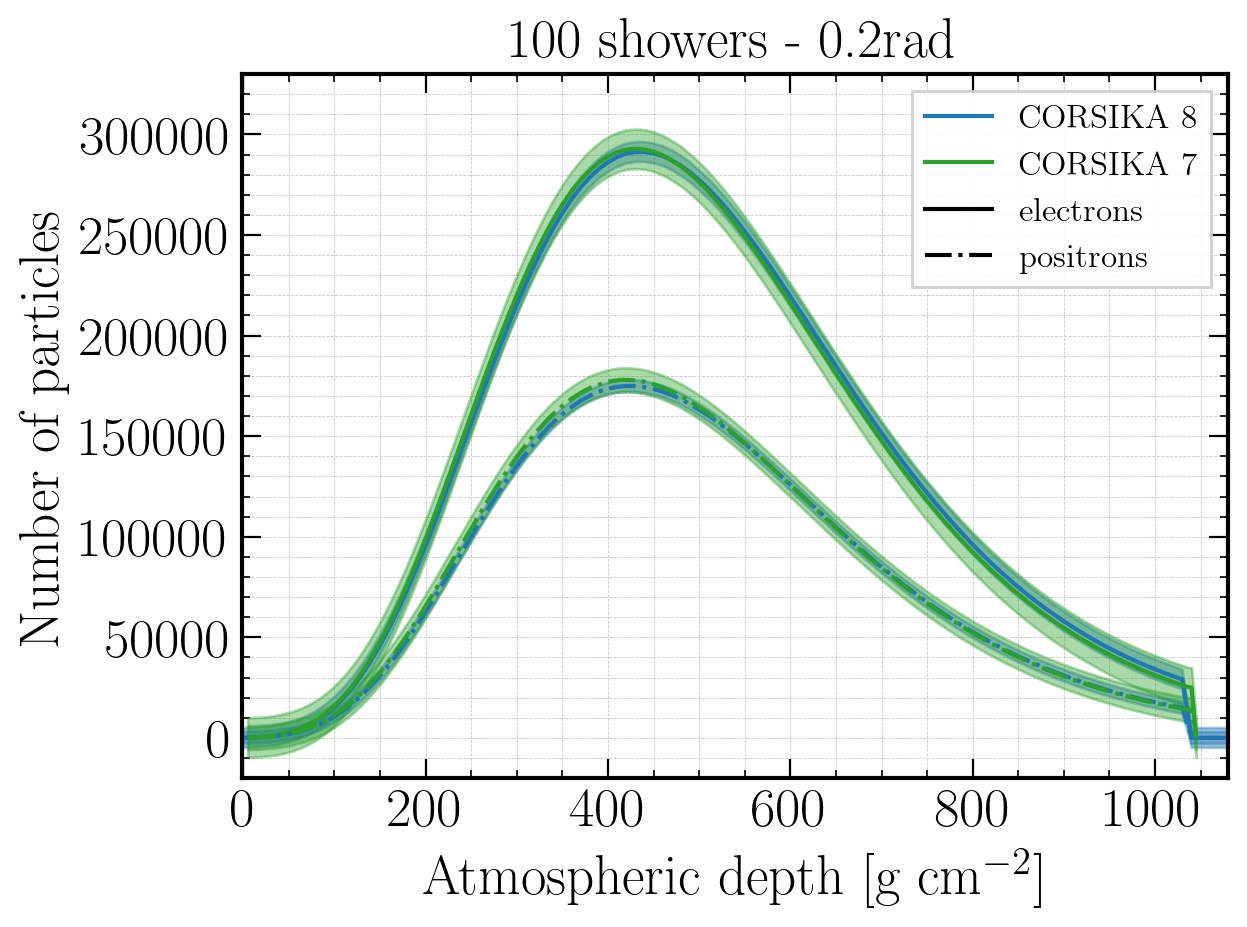}
\end{subfigure}%
\begin{subfigure}{.5\textwidth}
    \centering
    \includegraphics[width=7.5cm]{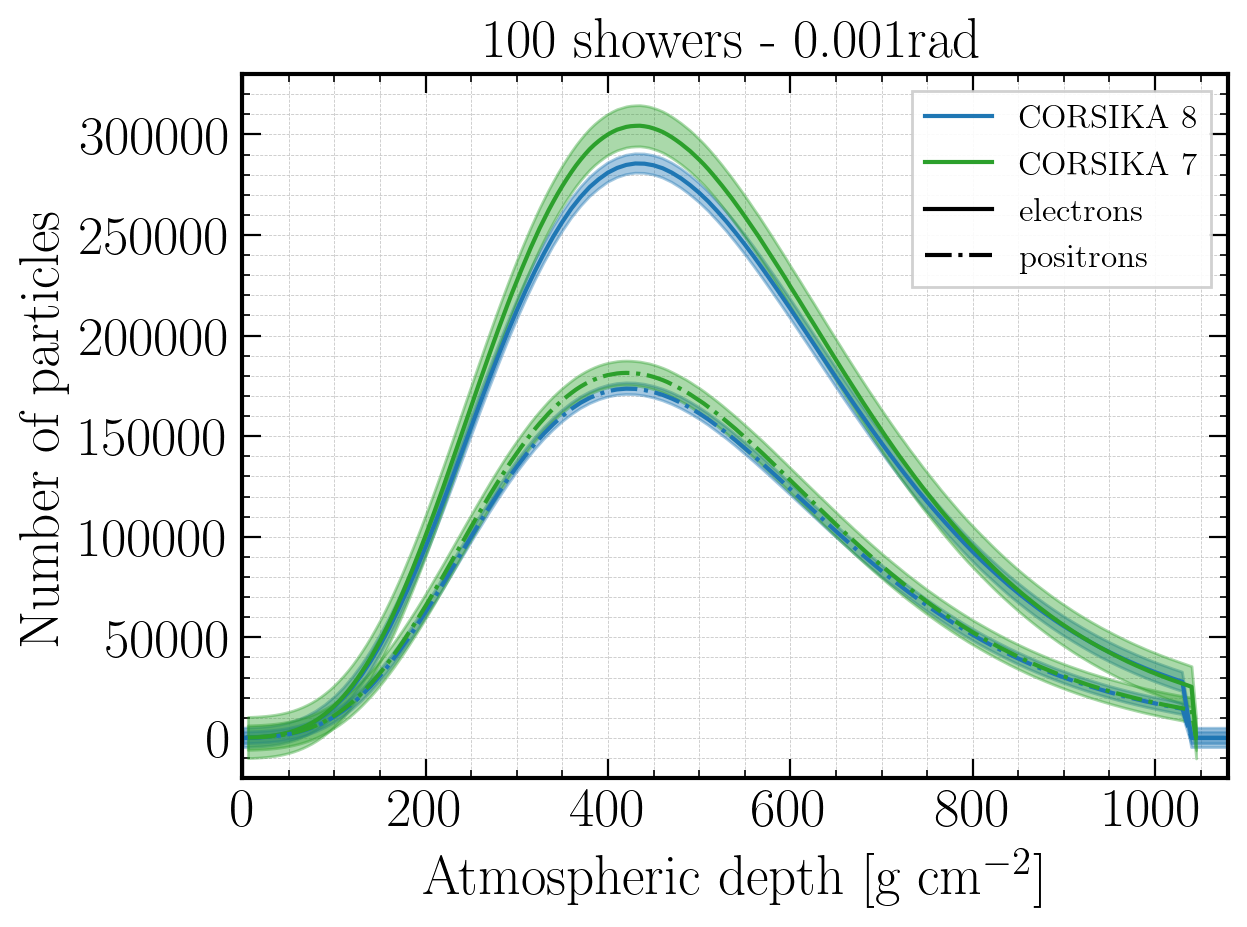}
\end{subfigure}
\caption{Mean longitudinal profiles along with their statistical uncertainty band (error of the mean) for 100 showers with the maximally allowed magnetic deflection angle set to 0.2 rad (left) and to 0.001 rad (right). The sudden drop to zero at 1050 g/cm$^2$ occurs because the shower reaches the ground.}
\label{fig:c8vsc7-longprofs}
\end{figure*}

Now that we have established the relevance of choosing a sufficiently small step size for particle tracking within C8, we will perform a more detailed comparison between C8 and C7, investigating the impact of the step size on the longitudinal profiles, fluence footprint, and radiation energy. For C8 we use the simulations produced in \cref{sec:coreas-vs-zhs} with the ``CoREAS'' formalism. For C7, we produce simulations configured as close as possible to the C8 ones. Hence, we have 100 showers simulated with C7 where the MaxRad value is set to 0.2 rad and 100 more where MaxRad is set to 0.001 rad. In all simulations, the point of first interaction has been fixed in order to minimize shower-to-shower fluctuations.

The longitudinal profiles for each of the two MaxRad values are shown in \cref{fig:c8vsc7-longprofs}. The agreement is better for the 0.2 rad case, while for 0.001 rad C7 predicts more particles than C8, an effect that should be studied in more depth. For the comparison of the radio emission, we normalize the energy fluences (and radiation energies) by the square of the electromagnetic energy of the underlying showers to take out this systematic difference in the longitudinal profiles between C7 and C8.

\begin{figure*}
    \centering
    (($f$(C7) - $f$(C8)) / $f$(C7),  \SIrange{30}{80}{\mega\hertz}\\

    \vspace{1em} 

    \threeByTwoGrid
    {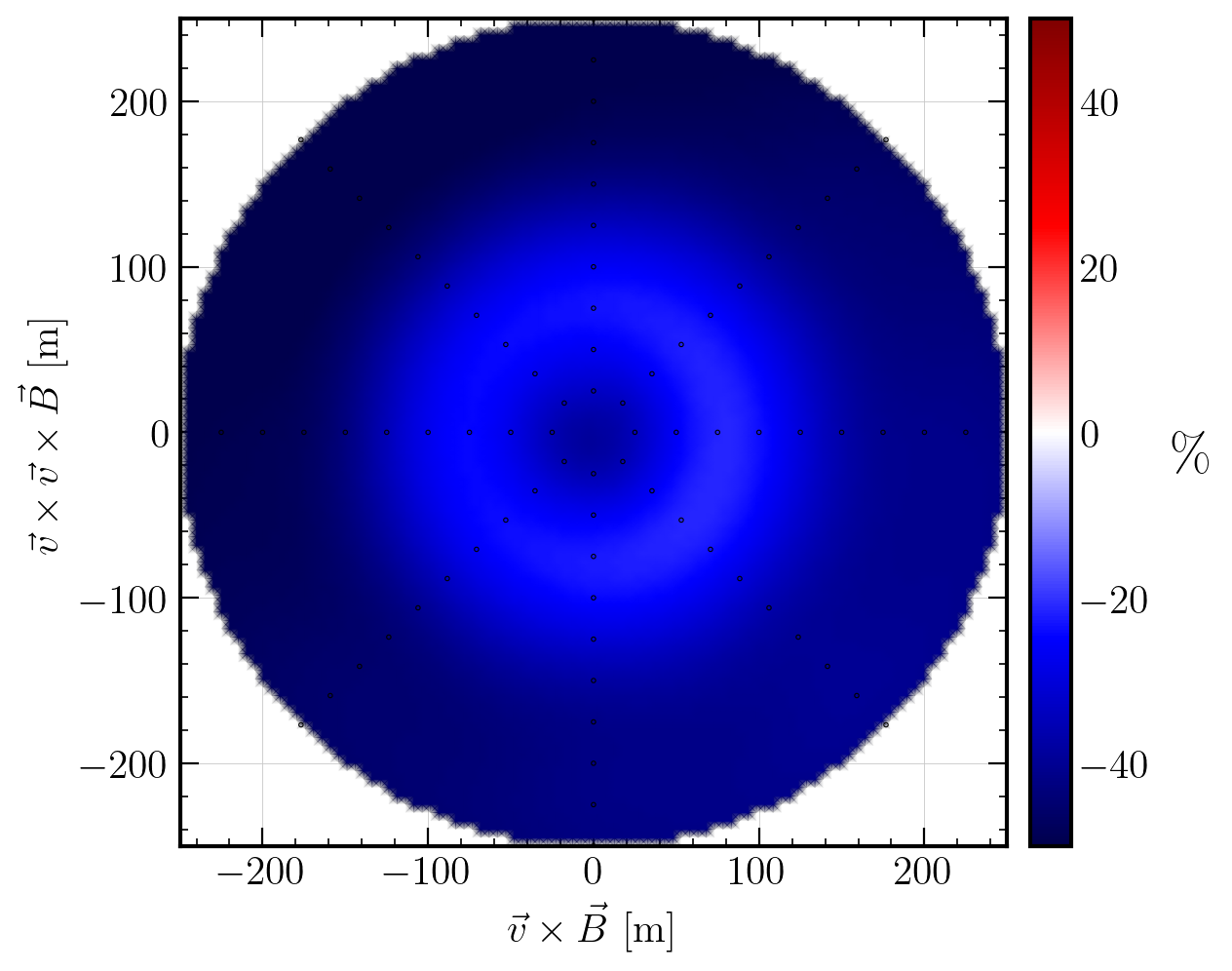}
    {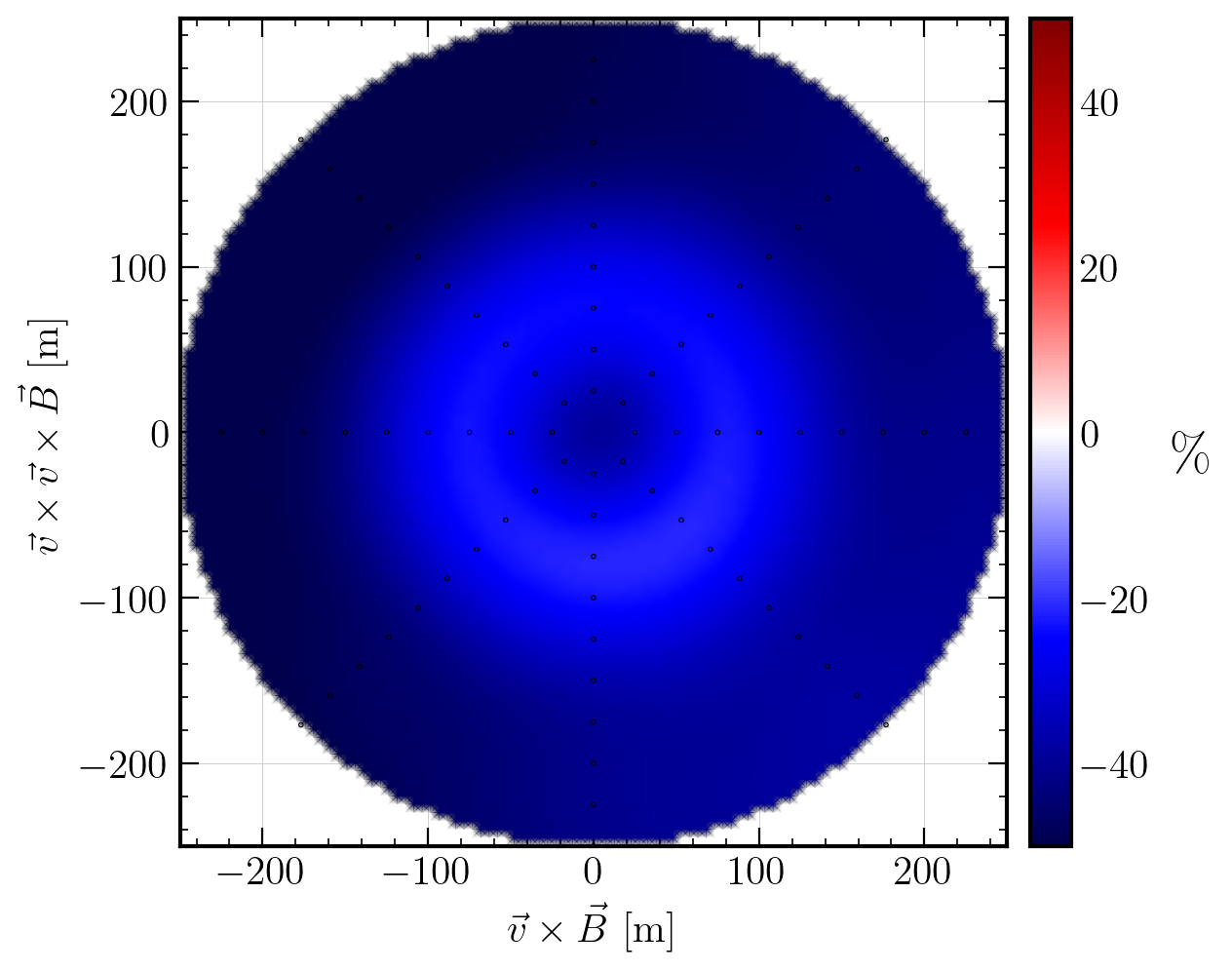}
    {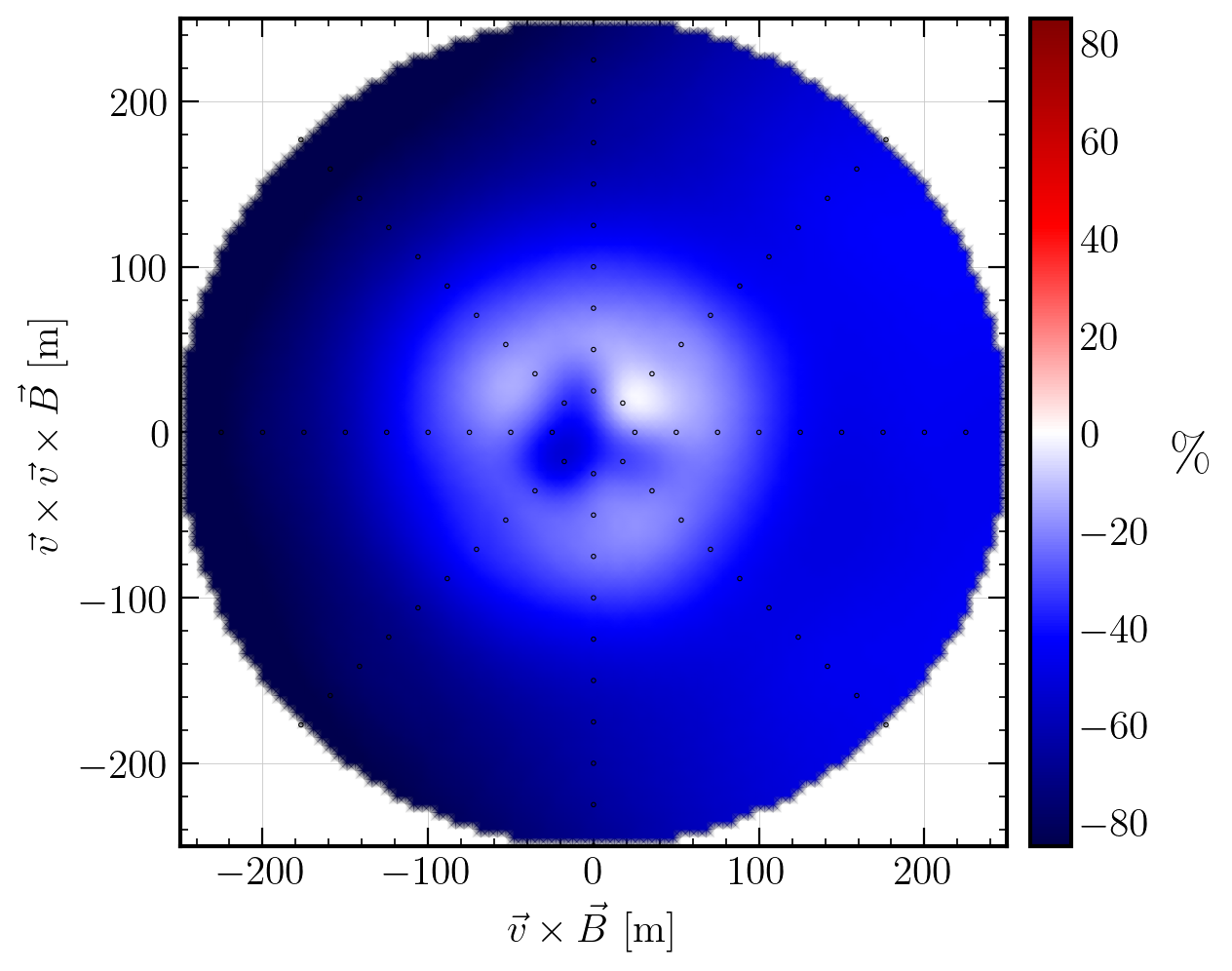}
    {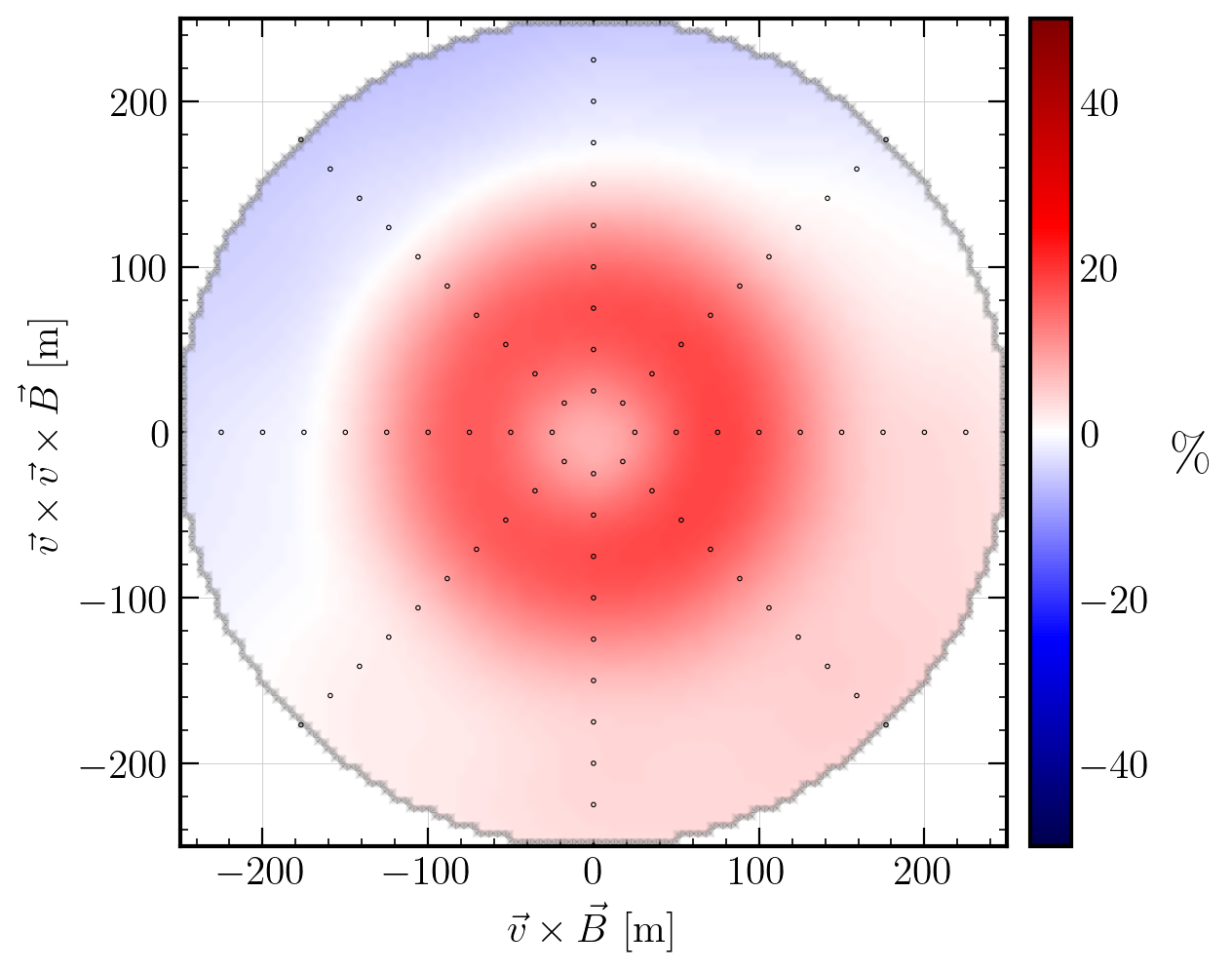}
    {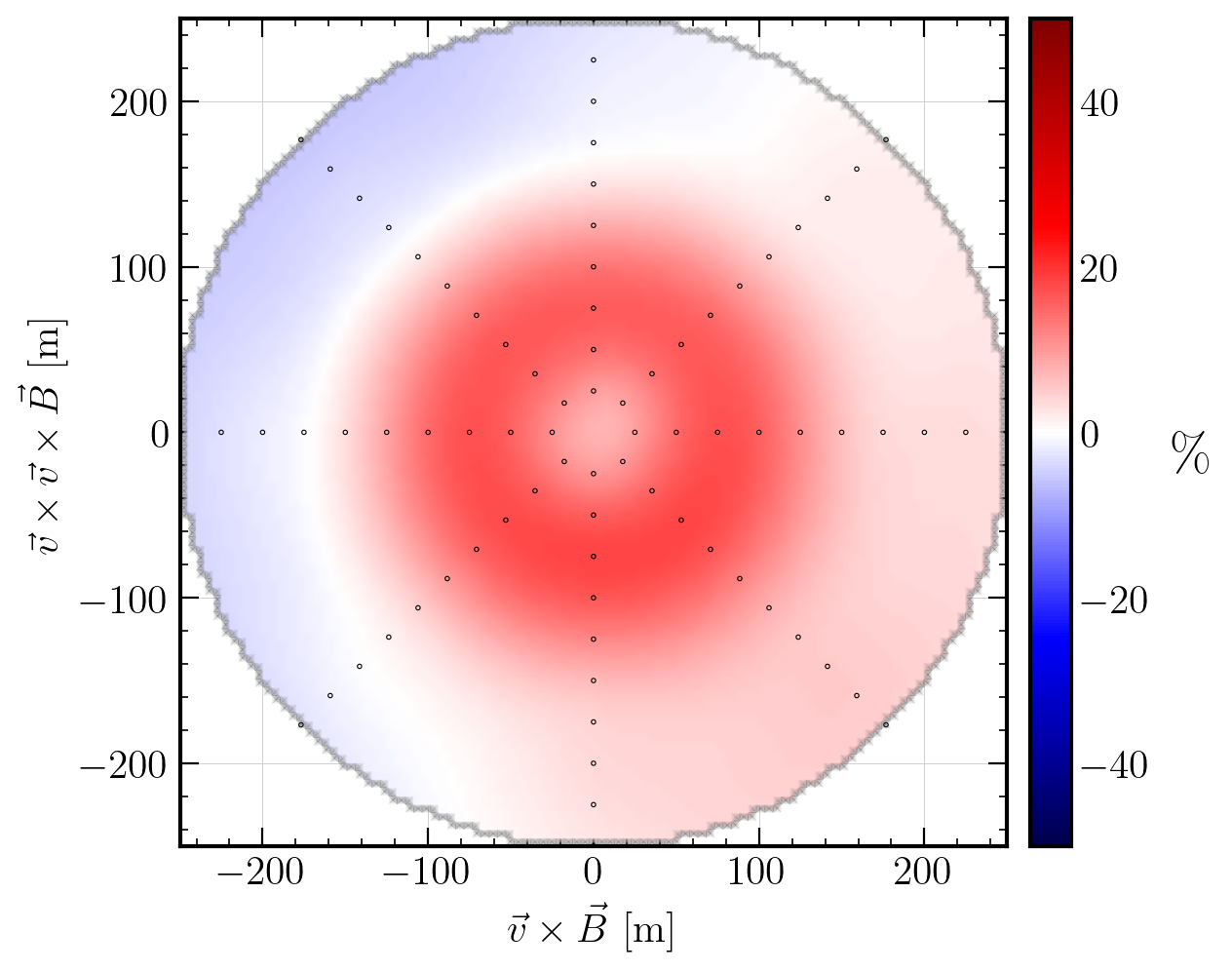}
    {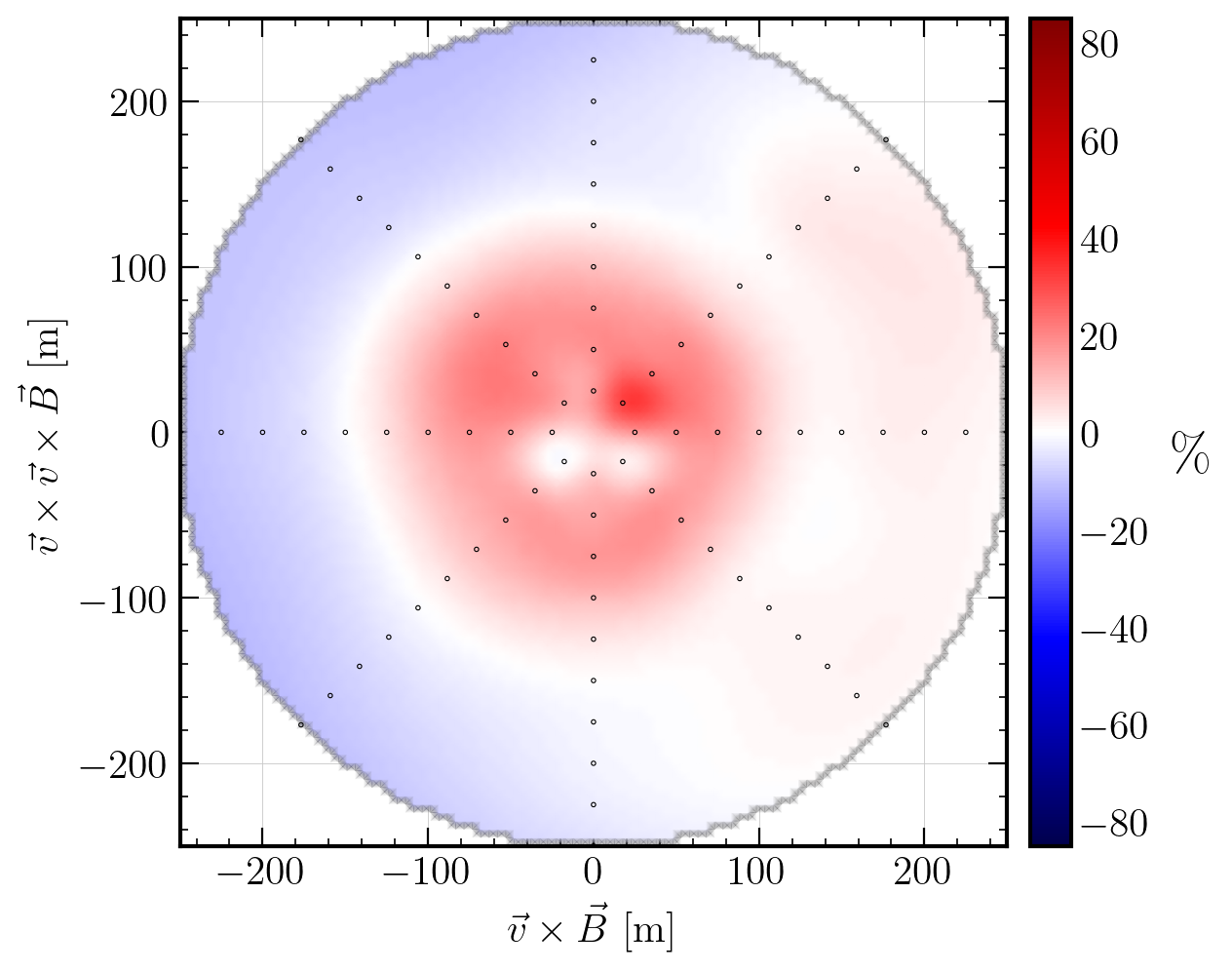}
    
\caption{Percentage deviation maps for the energy fluence simulated with C7 and C8 with ``CoREAS'' formalism averaged over 100 iron-induced vertical air showers with an energy of \SI{1}{\peta\eV} in the \SIrange{30}{80}{\mega\hertz} frequency band. Upper row is for MaxRad value of 0.2, lower row is for MaxRad value of 0.001. From left to right: total fluence, geomagnetic fluence, and charge-excess fluence.}
\label{fig:c8c7-02-0001-maxRad-30-80}
\end{figure*}

We use the same strategy for the fluence footprint comparisons as in \cref{sec:coreas-vs-zhs} and use the C7 simulations as reference here. In \cref{fig:c8c7-02-0001-maxRad-30-80} the ratio maps for total fluence, geomagnetic and charge excess contributions in the \SIrange{30}{80}{\mega\hertz} frequency band for the 0.2 rad (top) and 0.001 rad (bottom) cases are shown. Red color indicates that C7 predicts higher fluence, while blue color indicates that C8 predicts higher fluence. For the case of 0.2 rad, the differences are large, of the order of 30\%, as was already seen in section \ref{sec:air-shower-comparison}. C8 consistently predicts more energy fluence and hence, radiation energy. However, for the case of small particle tracks with a MaxRad value of 0.001, in the same frequency band, the agreement is much better, on the level of 10\% for the 30-80\,MHz band.

\begin{figure*}
    \centering
    (($f$(C7) - $f$(C8)) / $f$(C7),  \SIrange{50}{350}{\mega\hertz}\\

    \vspace{1em} 

    \threeByTwoGrid
    {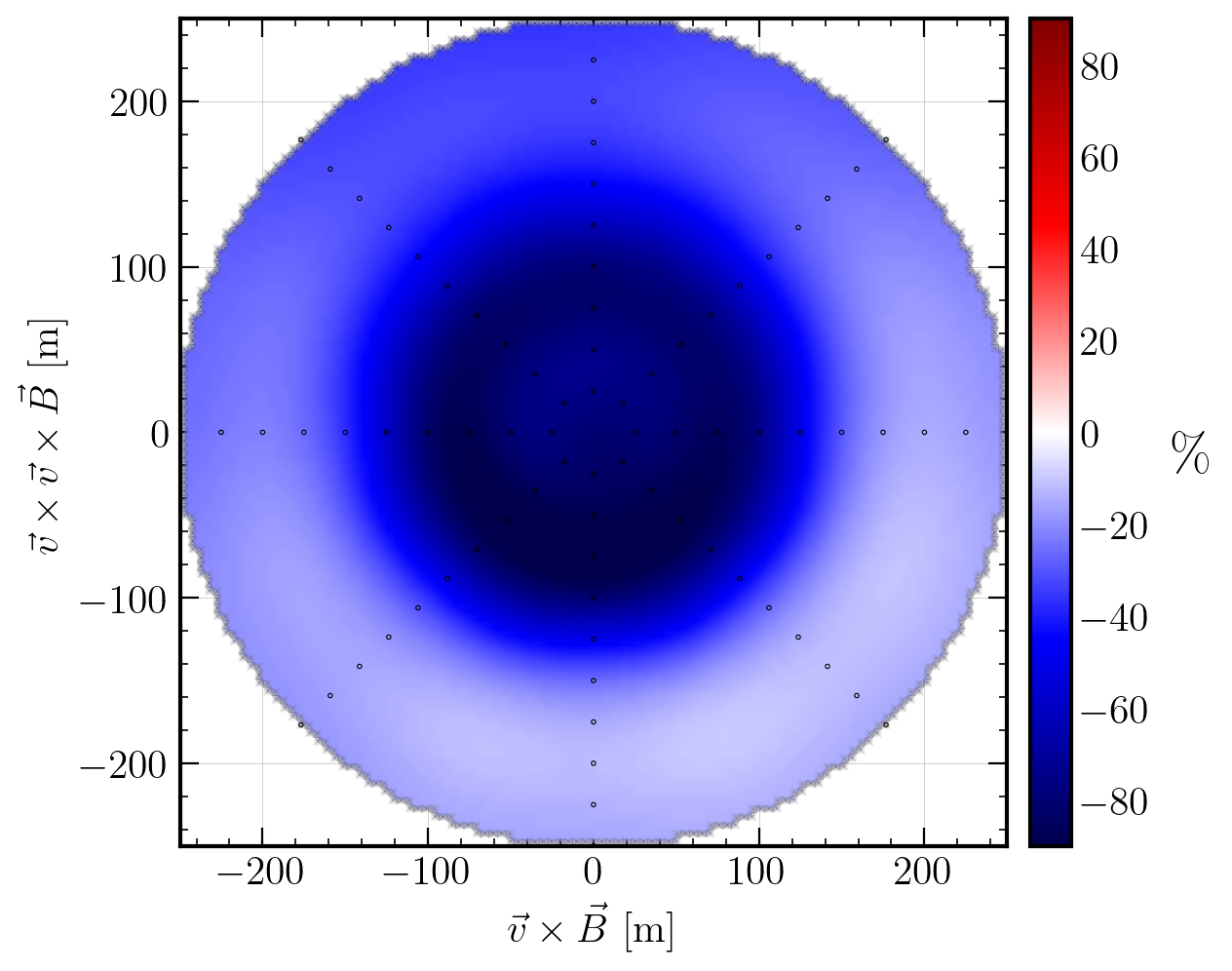}
    {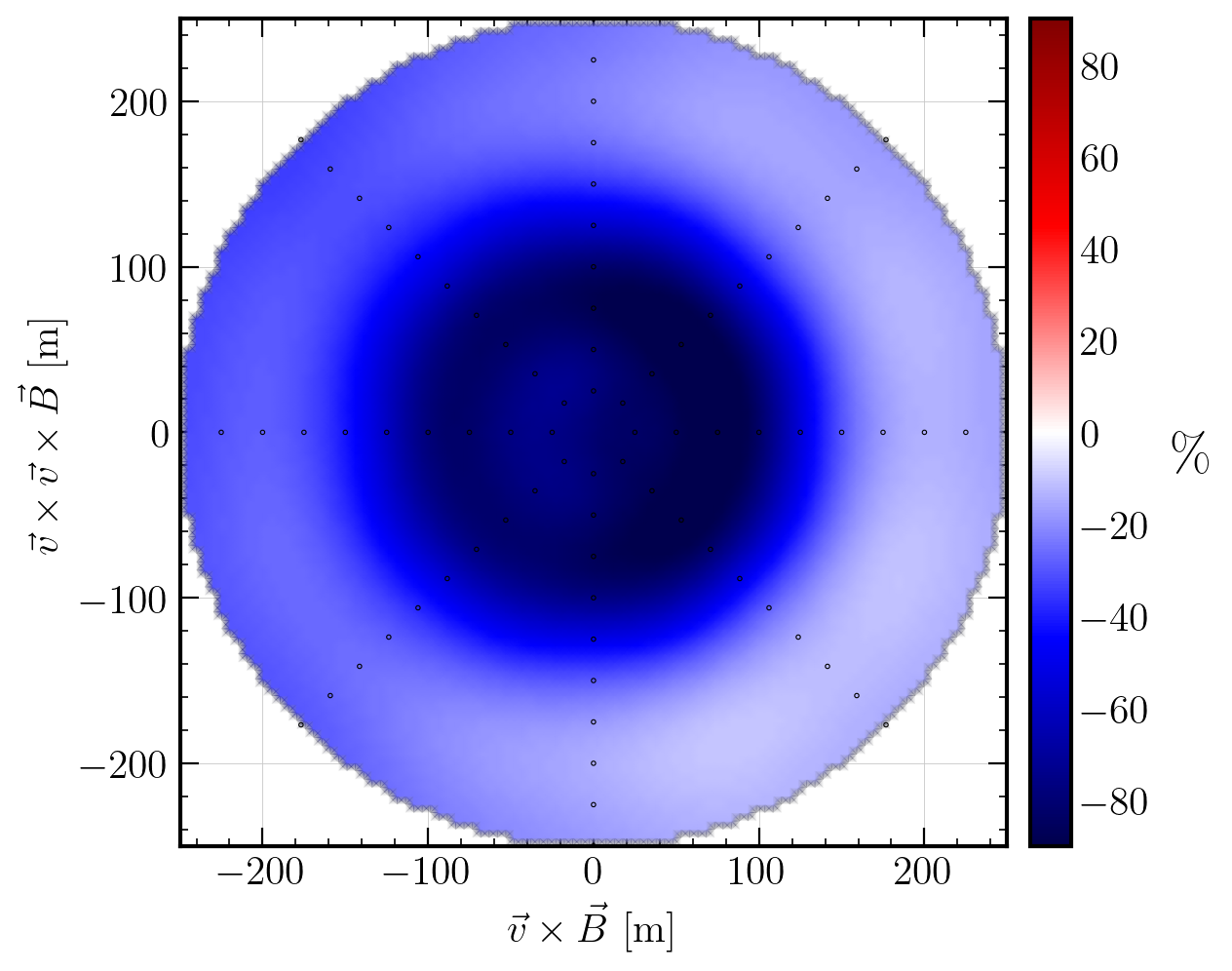}
    {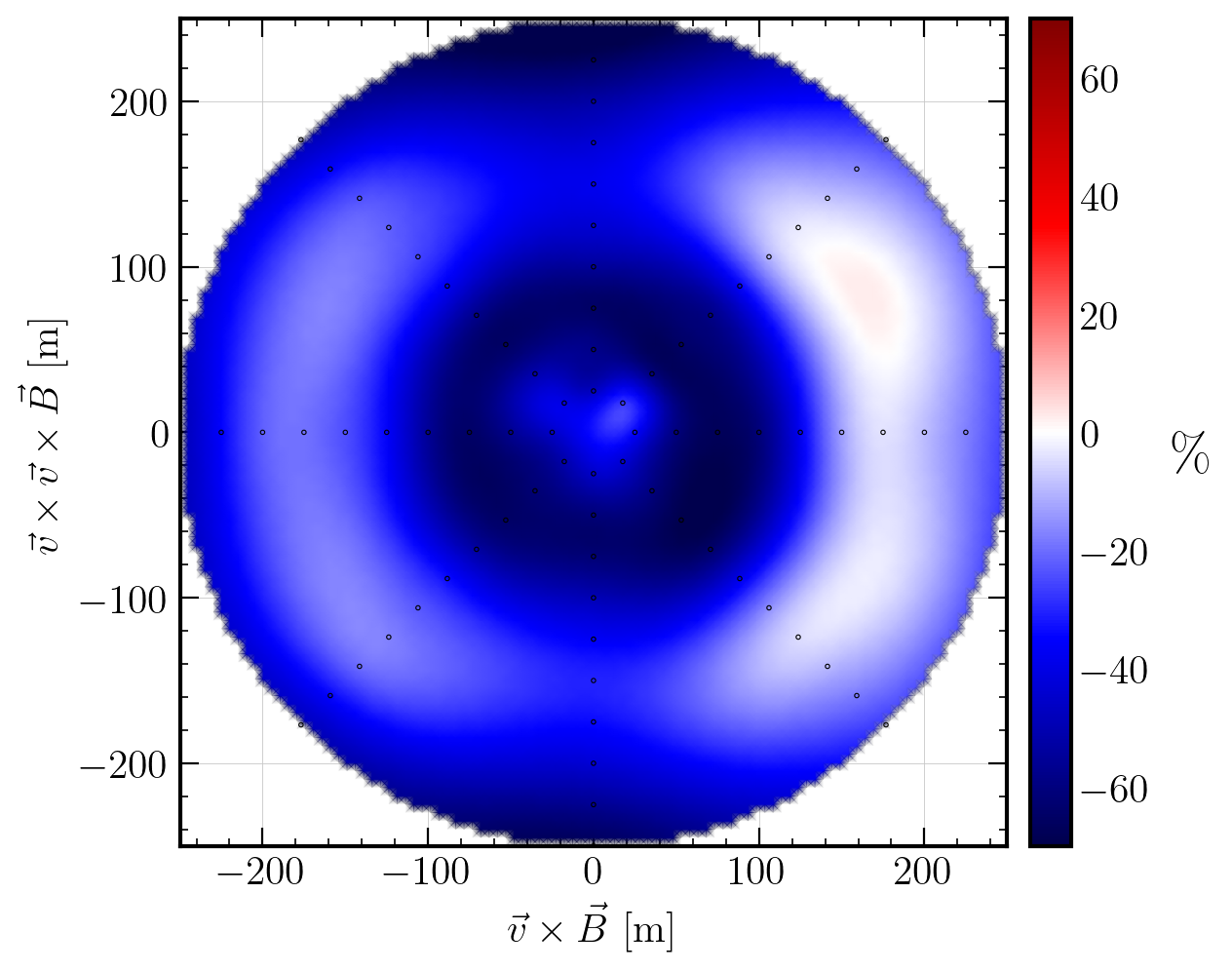}
    {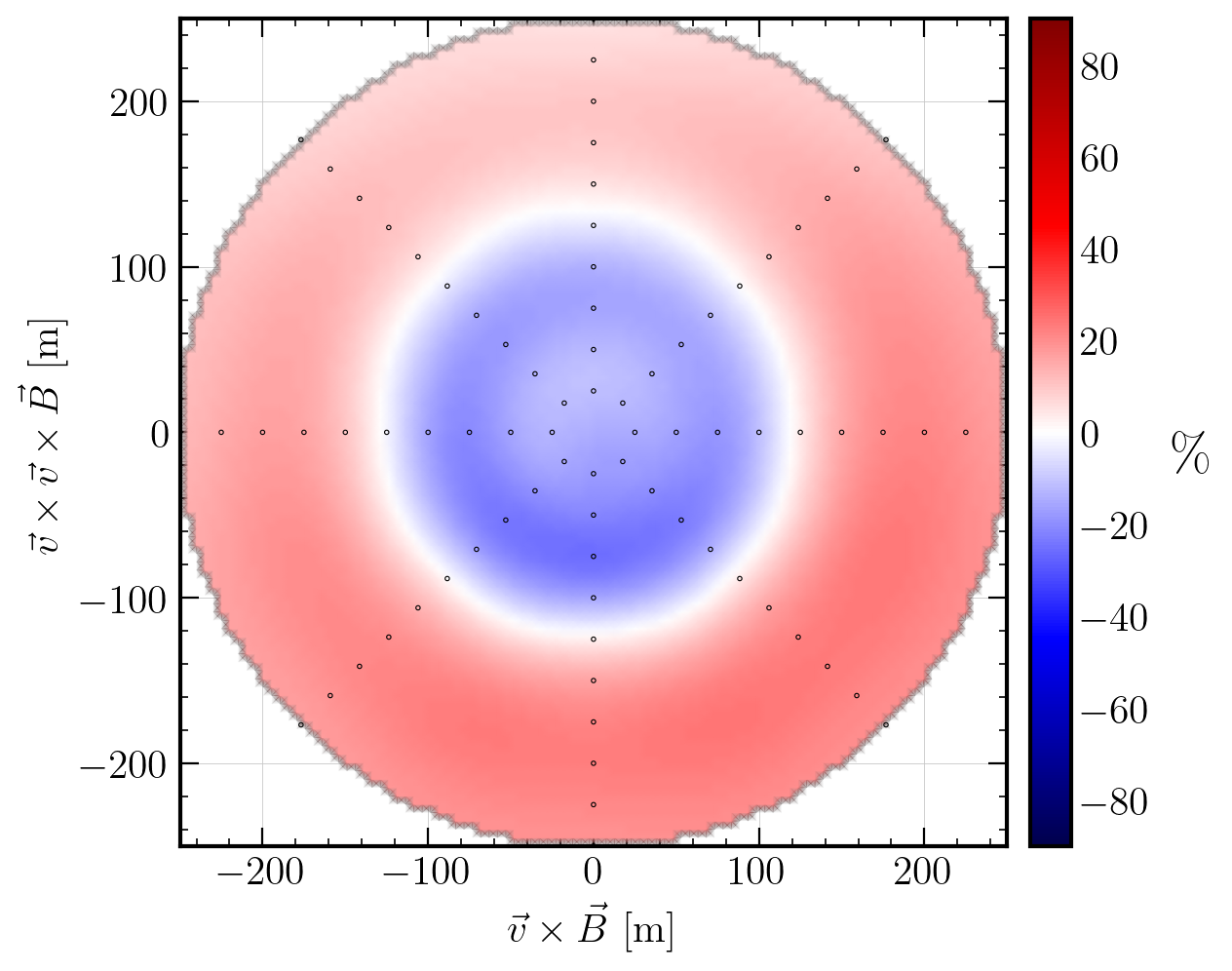}
    {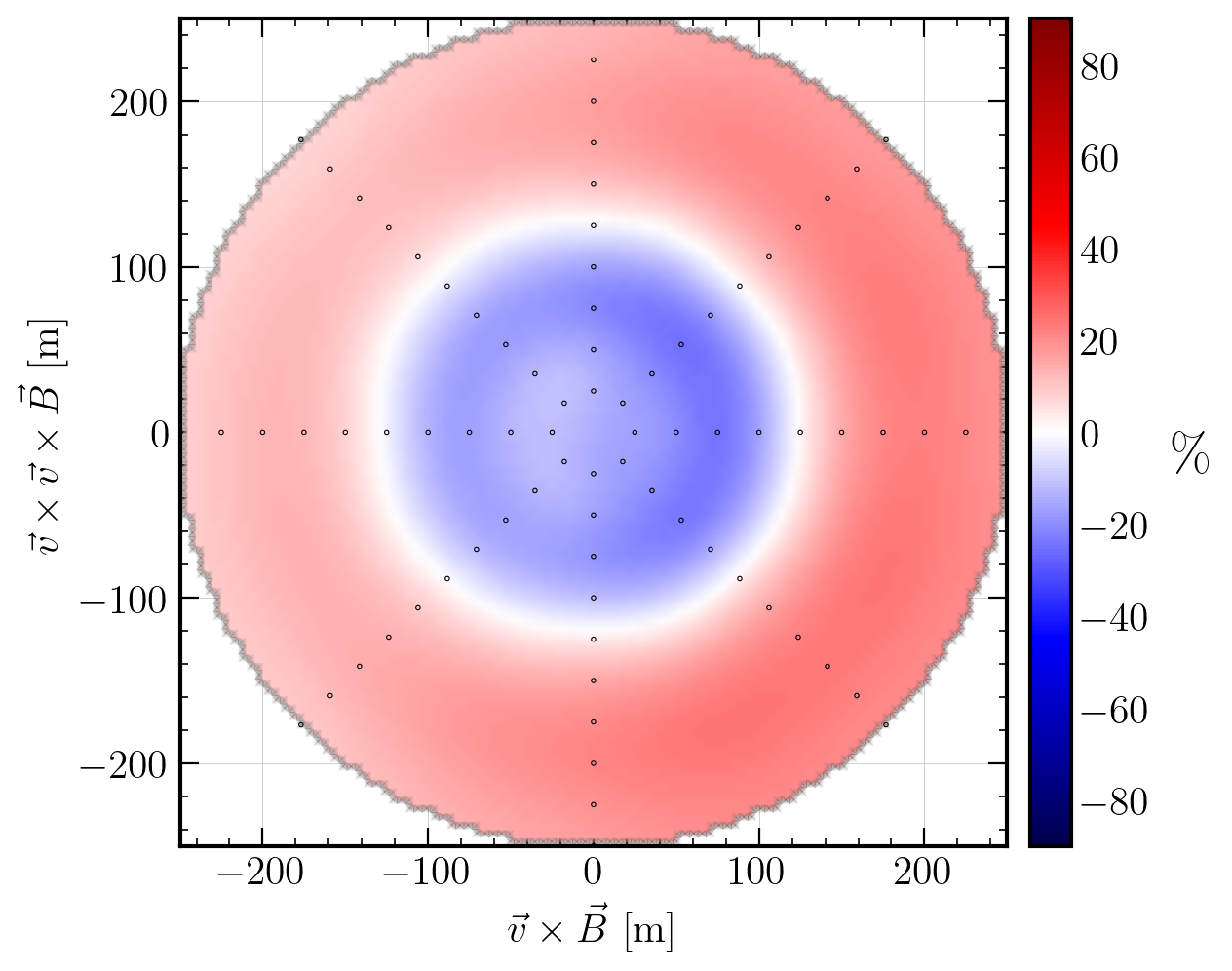}
    {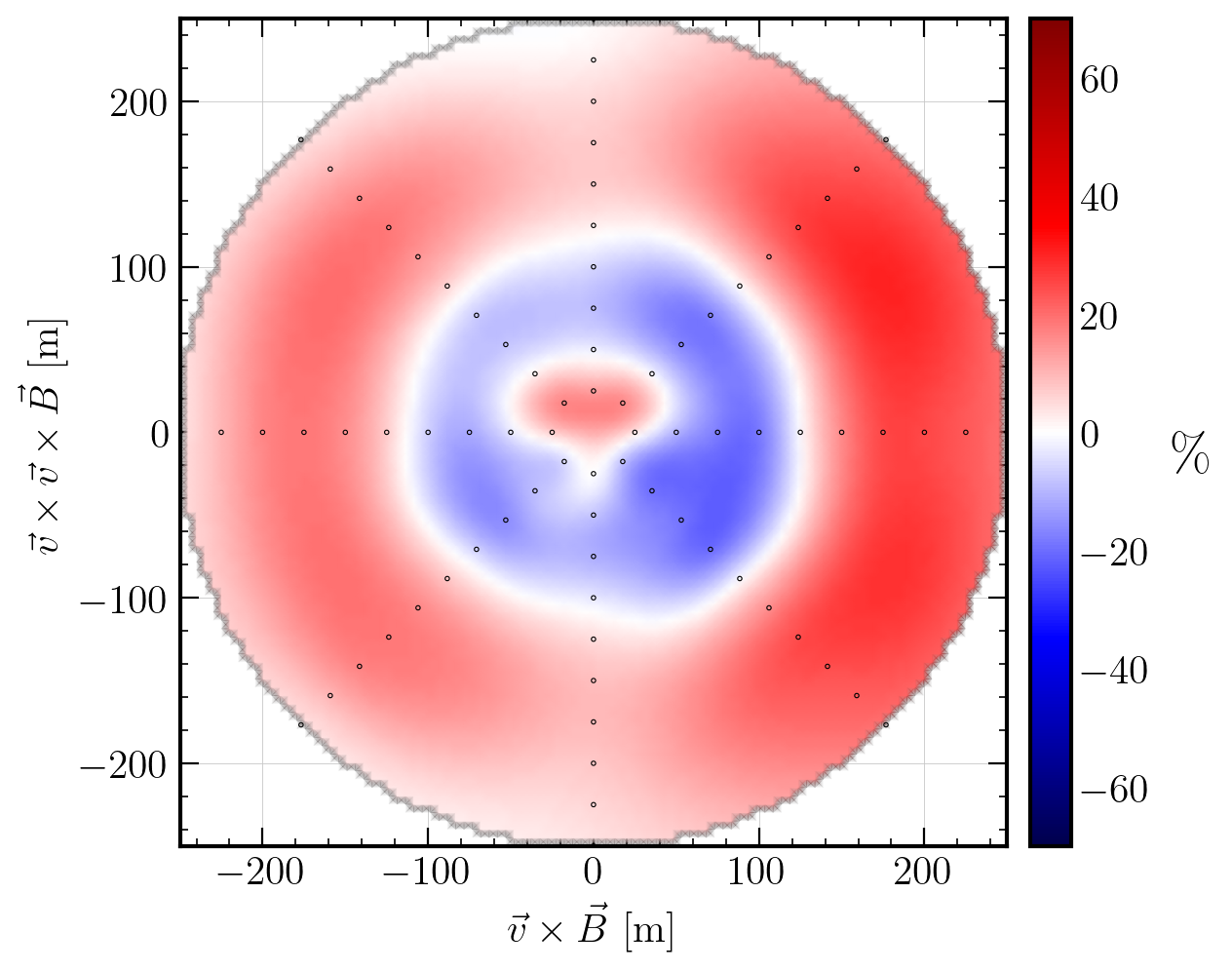}
    
\caption{Same as figure \ref{fig:c8c7-02-0001-maxRad-30-80} but for the 50-350\,MHz band.}
\label{fig:c8c7-02-0001-maxRad-50-350}
\end{figure*}

We find a similar trend for the \SIrange{50}{350}{\mega\hertz} frequency band as shown in \cref{fig:c8c7-02-0001-maxRad-50-350}. The agreement between the two codes is much better for shorter particle tracks. That said, even at small step sizes the fluence maps exhibit some structure, with C8 predicting $\sim 10\%$ higher fluence within the Cherenkov ring and $\sim 10\%$ lower fluence outside the Cherenkov ring. Again, this is likely related to differences in the lateral distributions of electrons and positrons predicted by the two codes and should be investigated in more depth in the future.

\begin{table}
\centering
\begin{tabular}{l 
                S[table-format=+2.1, table-space-text-post={\,\%}] 
                S[table-format=+2.1, table-space-text-post={\,\%}]}
\toprule
\textbf{Frequency Band} & \multicolumn{2}{c}{\textbf{MaxRad Value}} \\
\cmidrule(lr){2-3}
& \textbf{0.2 rad} & \textbf{0.001 rad} \\
\midrule
\textbf{30-80\,MHz} &   &  \\
Total Emission & -34.9\,\% & 8.3\,\% \\
Geomagnetic Emission & -34.8\,\% & 8.2\,\% \\
Charge Excess Emission & -49.3\,\% & 2.6\,\% \\
\midrule
\textbf{50-350\,MHz} &   &   \\
Total Emission & -57.1\,\% & -1.7\,\% \\
Geomagnetic Emission & -57.0\,\% & -1.8\,\% \\
Charge Excess Emission & -34.1\,\% & 6.0\,\% \\
\bottomrule
\end{tabular}
\caption{Percentage difference in radiation energy between C8 with ``CoREAS'' formalism and C7 simulations for different frequency bands and MaxRad settings. Positive values indicate more radiation energy from C7.}
\label{tbl:table-c7c8}
\end{table}

In table \ref{tbl:table-c7c8}, the predicted radiation energy is compared for the different frequency bands and MaxRad values for the ``CoREAS'' formalism.

\begin{figure*}
\centering
\includegraphics[width=0.45\textwidth]{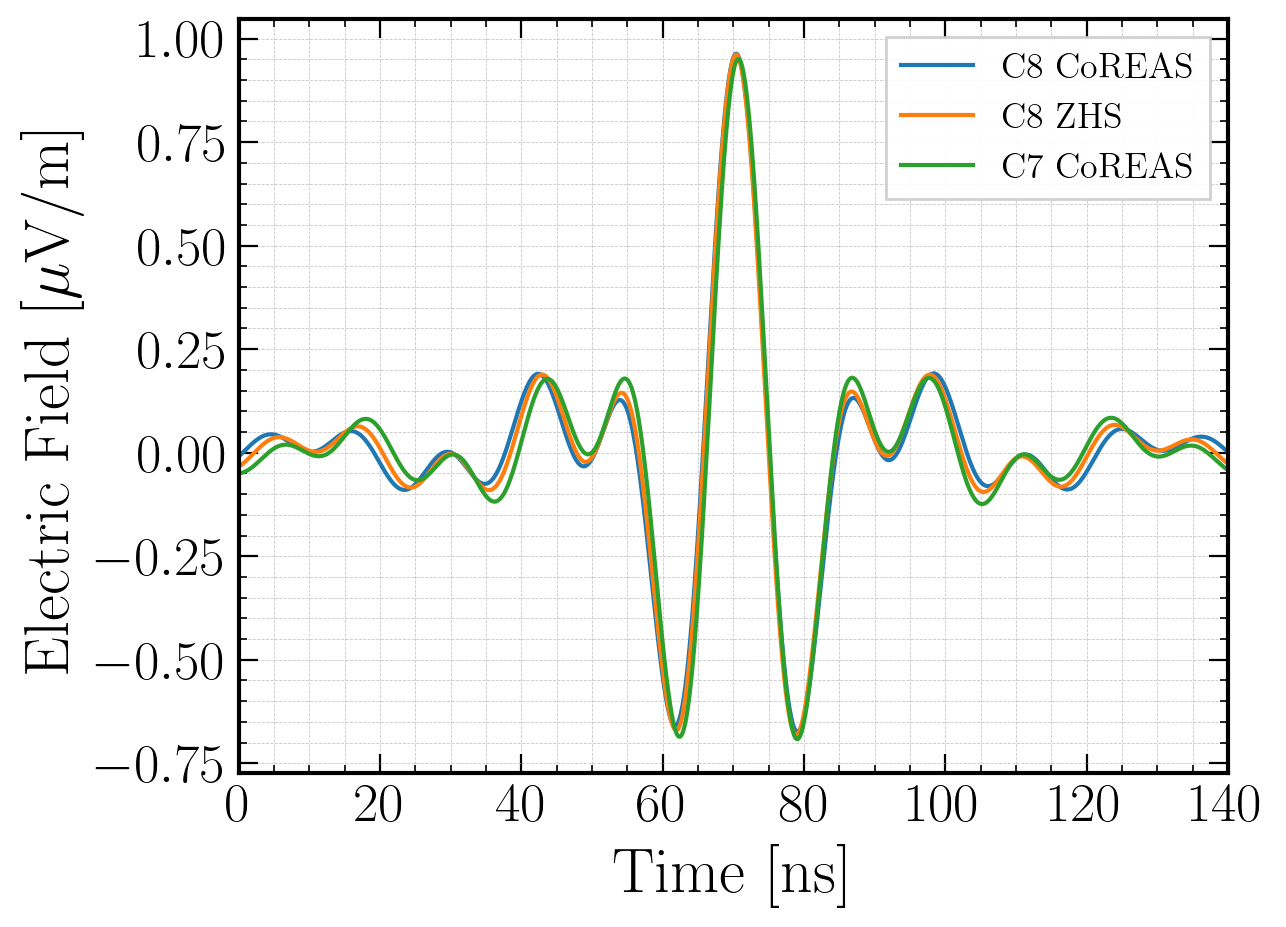}
\includegraphics[width=0.44\textwidth]{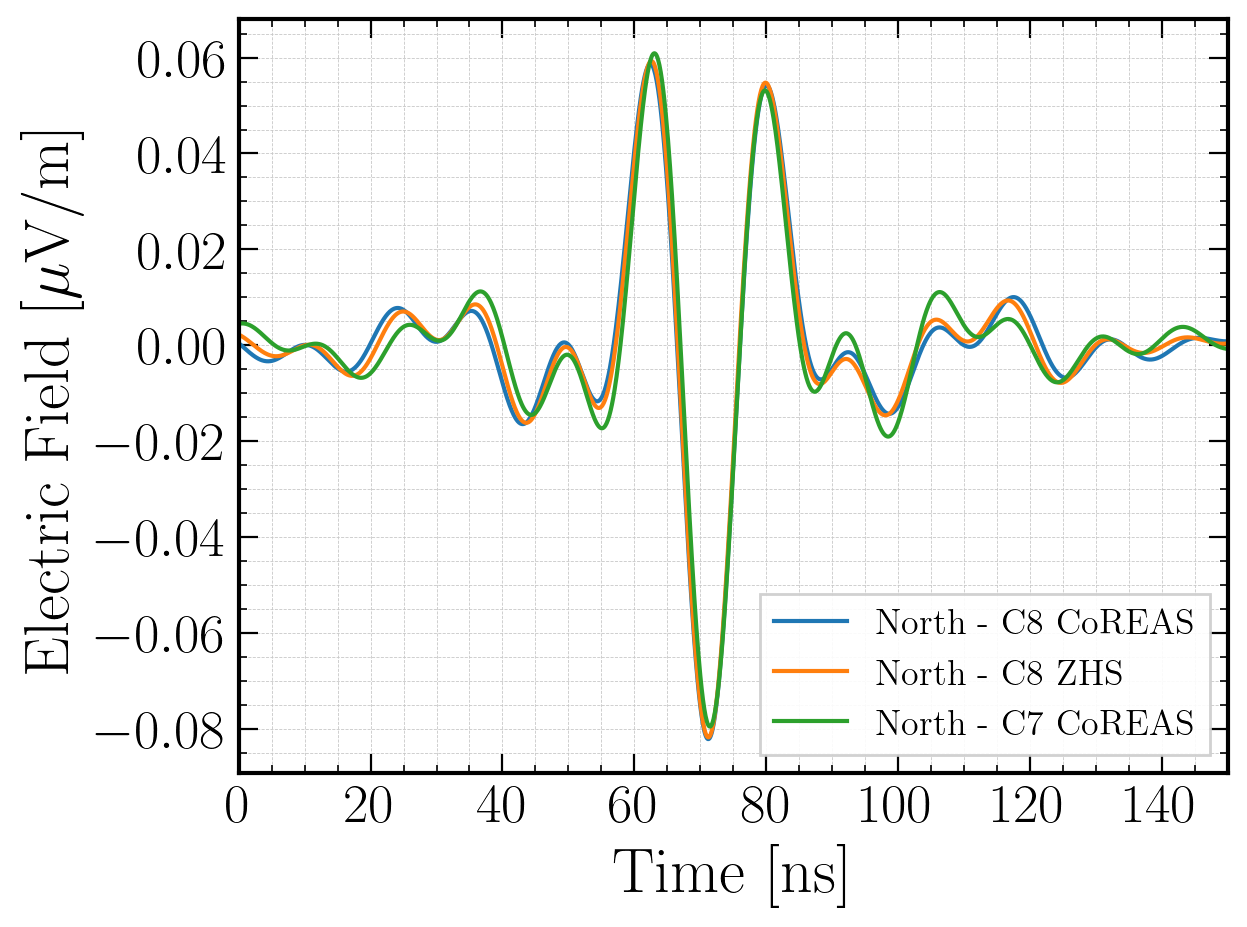}
\includegraphics[width=0.45\textwidth]{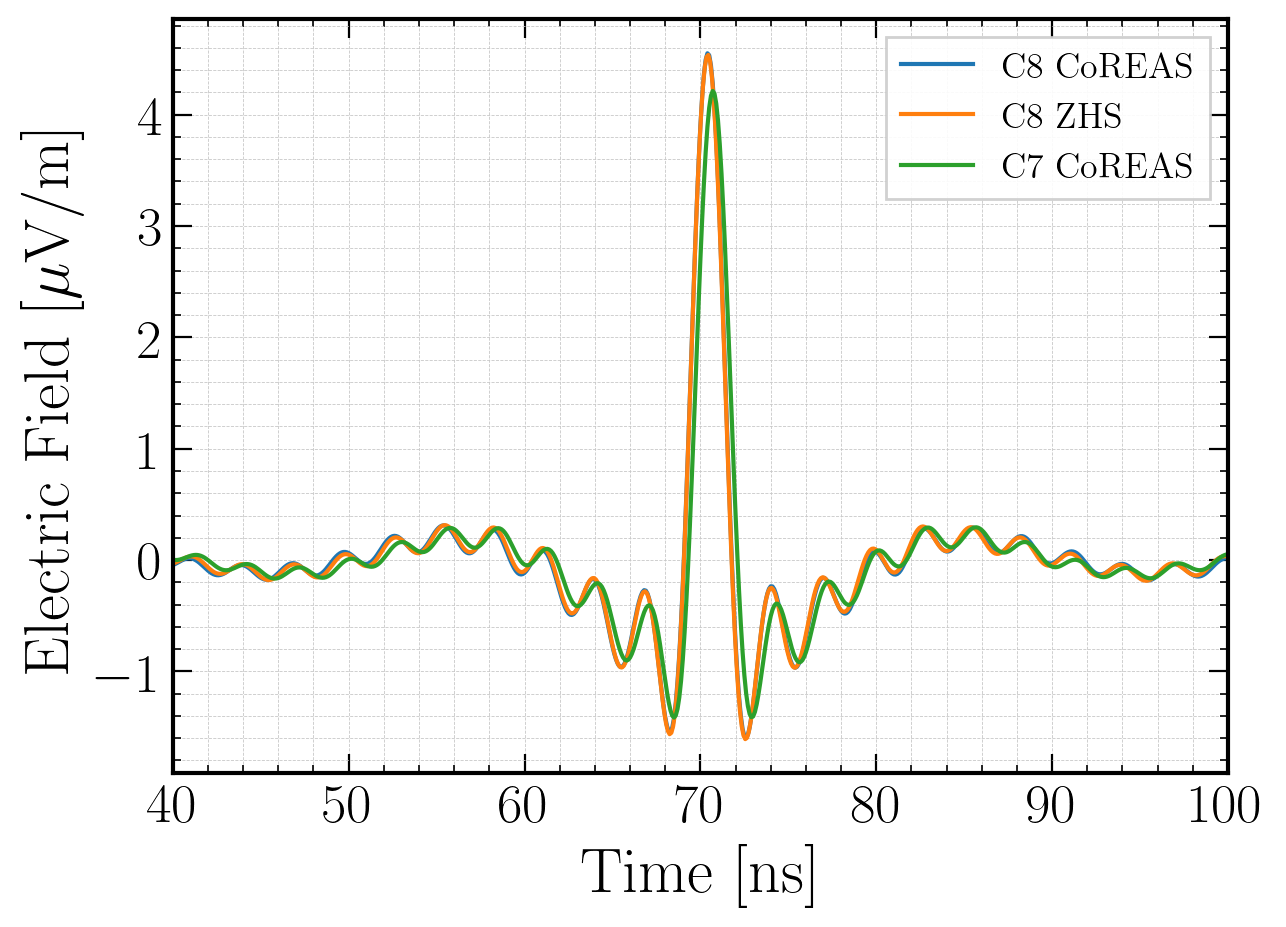}
\includegraphics[width=0.45\textwidth]{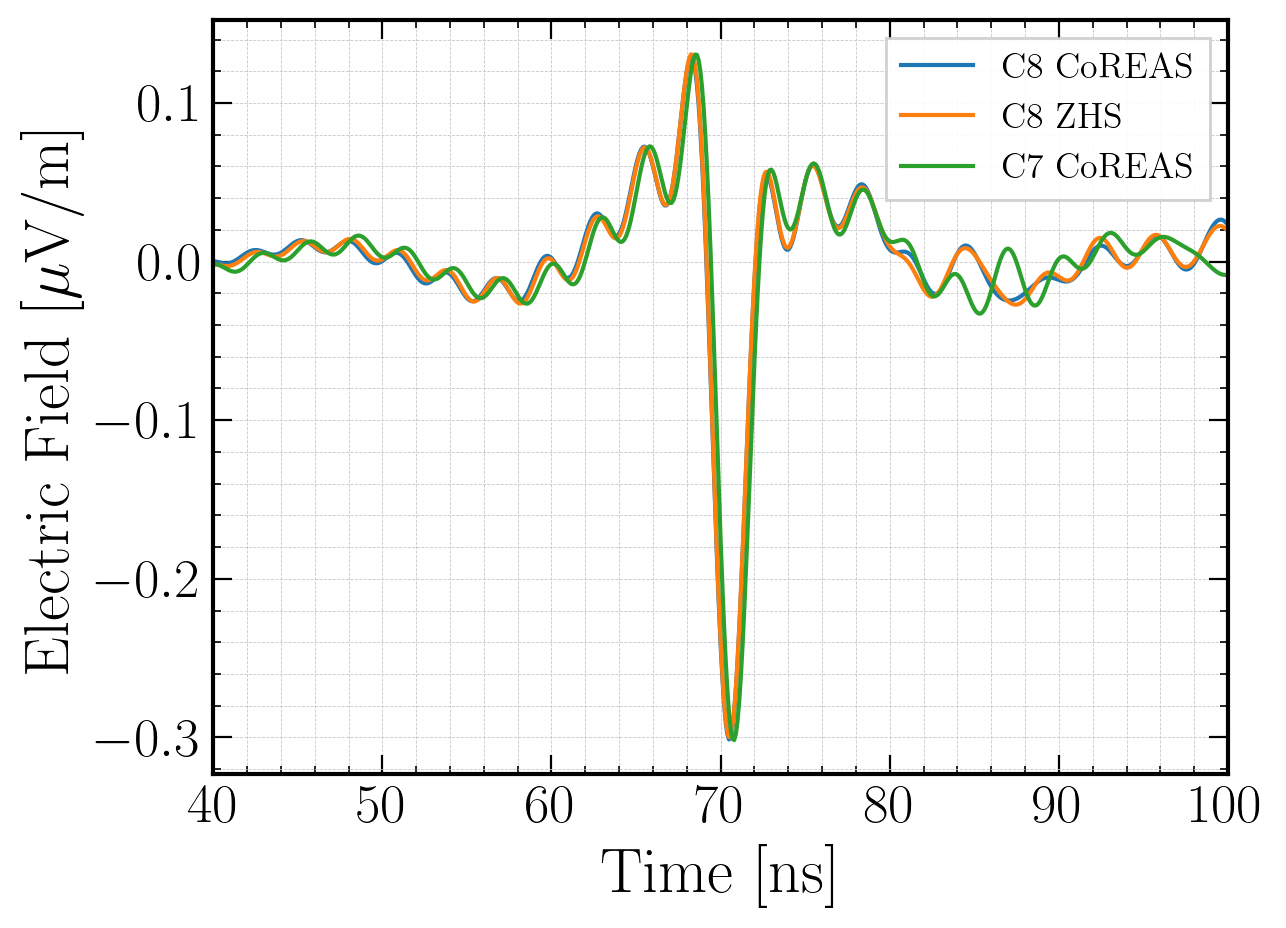}
\caption{Radio pulses predicted by C8 with ``CoREAS'' as well as ``ZHS'' formalisms and C7 for the geomagnetic emission (left) and charge-excess component (right) at a distance of 100\,m from the shower axis, i.e., close to the Cherenkov ring. Traces are filtered to the 30-80\,MHz band (top) and the 50-350\,MHz band (bottom).}
\label{fig:c8vsc7-pulses}
\end{figure*}

Finally, we select showers with very similar longitudinal profiles to explicitly illustrate the agreement of the radio pulses predicted by C8 with both formalisms and C7 in the 30-80\,MHz and 50-350\,MHz bands, respectively. Figure \ref{fig:c8vsc7-pulses} shows the predicted pulses for observers at an axis distance of \SI{100}{\metre}, i.e. on the Cherenkov ring, for a MaxRad setting of 0.001 rad in both codes. The agreement in both the 30-80\,MHz and 50-350\,MHz bands is at the percent-level.

\section{Influence of track length on C7 simulations} 
\label{sec:c7vsc7}

\begin{figure*}
  \centering
  (($f$(0.2 rad) - $f$(0.001 rad)) / $f$(0.2 rad), C7 CoREAS \\

  \includegraphics[width=0.45\textwidth]{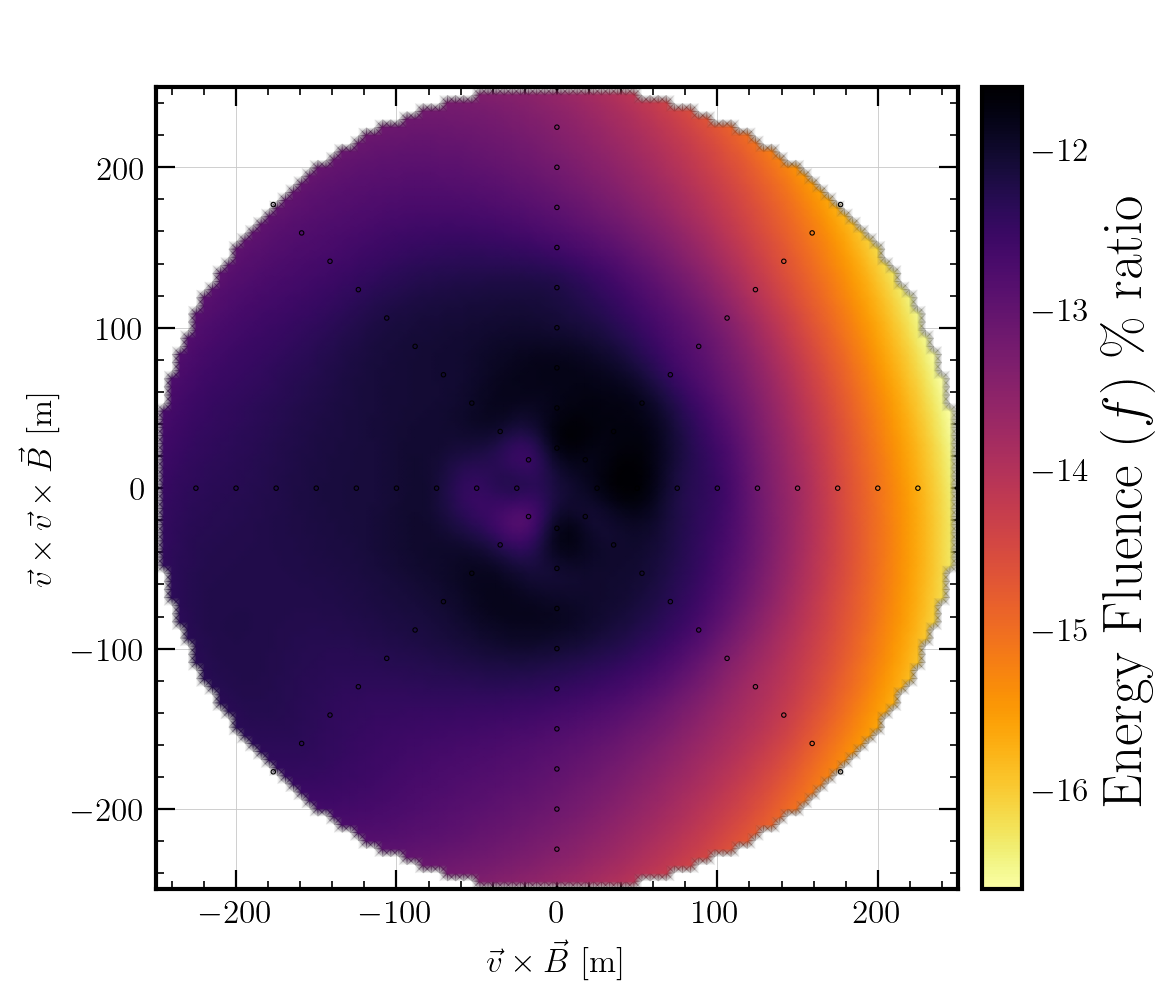}
  \hfill
  \includegraphics[width=0.45\textwidth]{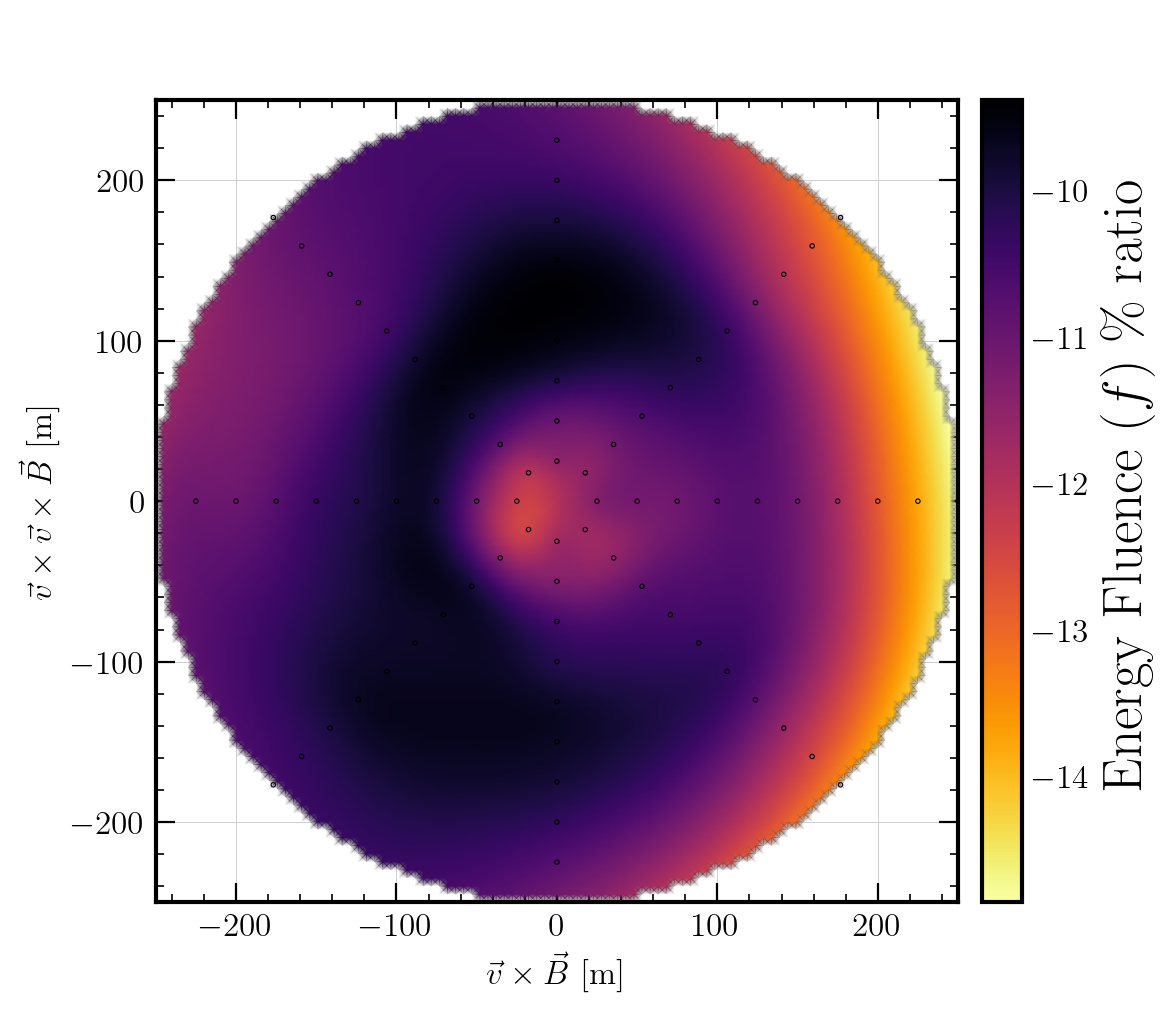}
\caption{Percentage deviation maps for the energy fluence simulated with C7 with default MaxRad value of 0.2 versus a MaxRad value of 0.001, averaged over 100 iron-induced vertical air showers with an energy of \SI{1}{\peta\eV} in the \SIrange{30}{80}{\mega\hertz} frequency band (left) and the \SIrange{50}{350}{\mega\hertz} frequency band (right). The negative values on the scales indicate that the simulations performed with the 0.001 rad setting produce higher fluence.}
\label{fig:c7vsc7-total-fluence}
\end{figure*}

It is also interesting to compare the differences in the fluence distributions arising in C7 when the track length is reduced, for otherwise the same settings as in the previous section. This is illustrated in \cref{fig:c7vsc7-total-fluence}. The fluence and hence radiation energy rises by about 12\% on average when the step size is decreased by changing the MaxRad parameter from 0.2 rad to 0.001. This increase is not global; some structure does arise in the footprint, albeit at the level of only roughly $\pm 3\%$. 

In an earlier study \cite{Gottowik:2017wio}, where the track length in C7 had been reduced by means of setting the ``STEPFC'' parameter governing the step size used for the determination of multiple scattering to a value of 0.05 instead of the default value of 1.0, 11\% more radiation energy was found to be predicted. While reducing the STEPFC parameter is not the same as reducing the MaxRad parameter, both methods lead to shorter step sizes in the simulation and apparently predict an increase in radiation energy by the approximately same amount.

\section{Conclusions}
\label{sec:sum}

We have presented the design philosophy for the implementation of radio-emission simulations in the new CORSIKA~8 particle shower simulation framework. The modular design, separating logical components such as the \textit{Formalism}, the \textit{Propagator}, and the \textit{Observer}, will provide the required flexibility to address more complex simulation scenarios than air showers. This is relevant in particular for simulations of radio emission in dense media such as antarctic ice, or even for the simulation of air showers, and their radio emission, crossing from air into dense media.

With low-level tests for the calculation of synchrotron pulses we have first verified that the radio-emission implementation yields the expected result, both when tracking particles manually in a magnetic field and when letting them be tracked by the C8 functionality.

A comparison between radio-emission predictions of C8 with its two currently available formalisms and the established C7-CoREAS and ZHAireS codes showed very good qualitative agreement in terms of the symmetry and polarization of energy fluence footprints, but significantly higher fluence predicted by C8 in comparison with C7 and ZHAireS at simulation default settings.

A detailed study revealed that the predictions by both of the formalisms included in C8, the ``CoREAS'' formalism (``Endpoints'' formalism with ZHS-like fallback near the Cherenkov angle) and the ``ZHS'' formalism, depend on the step size chosen in the particle simulation. For sufficiently short step sizes the two formalisms converge to almost identical results within 2\% of energy fluence. The effects of changing the step size are more pronounced in C8 than they are in C7, which likely points to room for improvement in how tracking and multiple scattering are currently handled in C8. For the current particle-tracking and multiple scattering approach in C8 we recommend a maxRad setting of 0.001 for accurate simulations of air-shower radio emission in the geomagnetic field.

Having established very good agreement between the two formalisms in C8 for sufficiently small step sizes we moved on to comparing the predictions of C8 with the ``CoREAS'' formalism and C7, for both default step sizes and small step sizes. While at default step sizes differences of 30-50\% in radiation energy become apparent, with C8 predicting more radiation energy than C7, for sufficiently small step sizes these differences reach a level of below 10\% in the 30-80\,MHz band and become negligible in the 50-350\,MHz band.

A difference of 10\% in radiation energy amounts to a difference of 5\% in an energy scale determined from radio-emission simulations (as particle energy is proportional to the square root of radiation energy). The fact that two independent codes, with independent electromagnetic interaction models, introduce a systematic uncertainty on the level of 5\% on the determination of the energy scale from radio-emission simulations -- which is smaller than current experimental uncertainties -- is a very reassuring result. That said, some structure is visible in the comparison of energy fluence footprints between C8 and C7, especially at high frequencies. This is most likely related to the lateral distribution of electrons and positrons in the shower, which influences the lateral distribution of the radio emission as well as the overall strength of the emission due to coherence effects. Further studies of these details need to be undertaken in the future, as the fluence distribution is relevant for the reconstruction of the depth of shower maximum from radio-emission simulations.

Finally, we compared the fluence footprints predicted by C7 for standard settings versus short tracks. Reducing the step size also has an effect in C7 (although on a significantly smaller level than in C8). The main effect is that $\sim 12$\% more radiation energy is predicted for small step sizes in C7, consistent with results of previous studies \cite{Gottowik:2017wio}. Structure in the fluence footprints is on the level of $2-3$\%, which is likely small enough as to not have practical relevance for the reconstruction of the depth of shower maxmum from radio-emission simulations.

In summary, CORSIKA~8 is ready for radio-emission simulations from air showers and particle showers in dense media. It has been successfully validated against existing codes, and we have demonstrated once more the robustness of microscopic radio-emission simulations across several different simulation codes and emission formalisms. In the future, details of the tracking algorithm in C8 will be further investigated to reduce the influence of the step size in simulations to the same level or better than in C7.

\section*{Acknowledgements}
This research was funded by the Deutsche Forschungsgemeinschaft (DFG, German Research Foundation) – Projektnummer 445154105 and Collaborative Research Center SFB1491 "Cosmic Interacting Matters - From Source to Signal". The authors acknowledge support by the High Performance and Cloud Computing Group at the Zentrum für Datenverarbeitung of the University of Tübingen, the state of Baden-Württemberg through bwHPC and the German Research Foundation (DFG) through grant no INST 37/935-1 FUGG. This research has been partially funded by the Federal Ministry of Education and research of Germany and the state of North Rhine-Westphalia as part of the Lamarr Institute for Machine Learning and Artificial Intelligence. This work has also received financial support from Ministerio de Ciencia e Innovaci\'on/Agencia Estatal de Investigaci\'on (PRE2020-092276). AC is supported by the Swedish Research Council {\sc (Vetenskapsrådet)} under project no.~2021-05449. CG is supported by the Swedish Research Council {\sc (Vetenskapsrådet)} under project no.~2021-05449 and the European Union under project NuRadioOpt/101116890. We also acknowledge support through UNAM PAPIIT project IN114924 and from the European Union’s Horizon 2020 research and innovation programme under the Marie Skłodowska-Curie grant agreement No. 101065027. The authors would further like to acknowledge fruitful discussion with Konrad Bernlöhr, Fan Hu, Paola Sala, the FLUKA collaboration.

\section*{CRediT author statement}
\sloppy

\textbf{Conceptualization}: Ammerman-Yebra, Arrabito, Alves Jr., Coleman, Huege, Karastathis, Reininghaus, Rhode, Riehn
\textbf{Methodology}: Alameddine, Ammerman-Yebra, Alves Jr., Coleman, Dembinski, Huege, Karastathis, Reininghaus, Rhode, Sandrock
\textbf{Software}: Alameddine, Ammerman-Yebra, Alves Jr., Coleman, Dembinski, Karastathis, Reininghaus, Riehn, Sampathkumar, Sandrock
\textbf{Validation}: Ammerman-Yebra, Coleman, Huege, Karastathis
\textbf{Formal analysis}: Gottowik, Karastathis
\textbf{Investigation}: Faure, Huege, Karastathis
\textbf{Resources}: Huege, Karastathis, Nellen, Reininghaus, Rhode
\textbf{Writing - Original Draft}: Huege, Karastathis
\textbf{Writing - Review \& Editing}: Alameddine, Alves Jr., Coleman, Gaudu, Glaser, Gottowik, Huege, Karastathis, Nellen, Reininghaus, Sandrock
\textbf{Visualization}: Gottowik, Karastathis
\textbf{Supervision}: Albrecht, Alves Jr., Elsässer, Engel, Glaser, Huege, Rhode
\textbf{Project administration}: Albrecht, Glaser, Huege, Nellen, Pierog, Reininghaus, Rhode, Sandrock
\textbf{Funding acquisition}: Albrecht, Elsässer, Engel, Glaser, Huege, Rhode






\end{document}